\documentclass[
reprint,
superscriptaddress,
amsmath,
amssymb,
aps,
pre,
showkeys,
floatfix,
]{revtex4-1}

\usepackage{bm}
\usepackage{comment}
\usepackage{graphicx}
\usepackage[
colorlinks=true,
citecolor=myBlue,
linkcolor=BrickRed,
urlcolor=blue,
breaklinks=true,
]{hyperref}
\usepackage{mathtools}
\usepackage{multirow}
\usepackage{subfigure}
\usepackage[normalem]{ulem}
\usepackage{wasysym}
\usepackage[usenames,dvipsnames]{xcolor}

\makeatletter
\def\@fnsymbol#1{
	\ensuremath{\ifcase#1\or
	\ddagger\or			
	*\or				
	\mathsection\or		
	\dagger\or			
	\mathparagraph\or	
	\|\or				
	**\or				
	\ddagger\ddagger\or	
	\dagger\dagger		
	\else\@ctrerr\fi}}
\makeatother

\allowdisplaybreaks

\definecolor{myBlue}{rgb}{0.0, 0.47, 0.75}

\newcommand{\bmx}{\bm{x}}
\newcommand{\bmv}{\bm{v}}

\newcommand{\kb}{k_{\mathrm{b}}}

\newcommand{\SL}{\mbox{\textit{S{\footnotesize{\hspace{-0.5pt}\textit{L}}}}}}	
\newcommand{\FvK}{F\"{o}ppl--von K\'{a}rm\'{a}n}

\newcommand{\Deltas}{\Delta_{\mathrm{s}}}

\newcommand{\sz}{{(0)}}

\newcommand{\tilh}{\tilde{h}}
\newcommand{\tilv}{\tilde{v}}
\newcommand{\tilr}{\tilde{r}}
\newcommand{\till}{\tilde{\lambda}}

\newcommand{\bfrac}[2]{{^{#1} \! / \! _{#2}}}

\newcommand{\nd}{*}
\newcommand{\ndl}{'}

\newcommand{\ellnd}{{\ell^{\nd}}}
\newcommand{\lambdand}{\lambda^{\nd}}

\newcommand{\phind}{\varphi^{\nd}}
\newcommand{\thetand}{\theta^{\nd}}
\newcommand{\phindl}{\varphi\ndl}
\newcommand{\thetandl}{\theta\ndl}
\newcommand{\tilhnd}{\tilh^\nd}
\newcommand{\tilrnd}{\tilr^\nd}
\newcommand{\tillnd}{\till^\nd}
\newcommand{\tnd}{t^\nd}
\newcommand{\xnd}{x^\nd}
\newcommand{\ynd}{y^\nd}
\newcommand{\znd}{z^\nd}
\newcommand{\xndl}{x\ndl}
\newcommand{\yndl}{y\ndl}
\newcommand{\zndl}{z\ndl}
\newcommand{\Deltand}{\Delta^{\! \nd}_{\mathrm{s}}}

\newcommand{\Deltandd}{\Delta^{\! \nd 2}_{\mathrm{s}}}

\begin{document}

%
%

\title{Geometry and dynamics of lipid membranes: The Scriven--Love number}

%
%

\author{Amaresh Sahu}
\email{amaresh.sahu@berkeley.edu}
\affiliation{Department of Chemical \& Biomolecular Engineering, University of California, Berkeley, CA 94720}

%
%

\author{Alec Glisman}
\email{alec.glisman@berkeley.edu}
\affiliation{Department of Chemical \& Biomolecular Engineering, University of California, Berkeley, CA 94720}

%
%

\author{Jo\"{e}l Tchoufag}
\email{jtchoufa@berkeley.edu}
\affiliation{Department of Chemical \& Biomolecular Engineering, University of California, Berkeley, CA 94720}

%
%

\author{Kranthi K. Mandadapu}
\email{kranthi@berkeley.edu}
\affiliation{Department of Chemical \& Biomolecular Engineering, University of California, Berkeley, CA 94720}
\affiliation{Chemical Sciences Division, Lawrence Berkeley National Laboratory, Berkeley, CA 94720}

\date{\today}

%
%

\begin{abstract}

The equations governing lipid membrane dynamics in planar, spherical,
and cylindrical geometries are presented here.
Unperturbed and first-order perturbed equations are determined and
non-dimensionalized.
In membrane systems with a nonzero base flow, perturbed in-plane and
out-of-plane quantities are found to vary over different length scales.
A new dimensionless number, named the Scriven--Love number, and
the well-known \FvK\ number result from a scaling analysis.
The Scriven--Love number compares out-of-plane forces arising from the in-plane,
intramembrane viscous stresses to the familiar elastic bending forces,
while the \FvK\ number compares tension to bending forces.
Both numbers are calculated in past experimental works, and span a wide
range of values in various biological processes across different geometries.
In situations with large Scriven--Love and \FvK\ numbers, the dynamical
response of a perturbed membrane is dominated by out-of-plane viscous
and surface tension forces---with bending forces playing a negligible role.
Calculations of non-negligible Scriven--Love numbers in various biological
processes and in vitro experiments show in-plane intramembrane viscous flows
cannot generally be ignored when analyzing lipid membrane behavior.

\end{abstract}

\maketitle

%
%

%
%

\section{Introduction} \label{sec:sec_intro}

Biological lipid membranes make up the boundary of the cell, as well as many of its internal organelles---including the nucleus, endoplasmic reticulum, and Golgi complex.
Such membranes are not simply static, semi-permeable barriers protecting their internal contents, but rather play a dynamic role in many cellular processes.
For example, at the neuronal synapse, spherical lipid membrane vesicles rapidly develop from planar membrane sheets to recycle lipids and proteins during ultrafast endocytosis \cite{watanabe-nature-2013}.
It is also known that thin membrane tubes can shoot suddenly from the endoplasmic reticulum into the cell cytoplasm, often fusing with one another when they cross~\cite{terasaki-jcb-1986, lippencott-science-2016}.
While there is much experimental evidence for the dynamic behavior of lipid membranes in biological systems, the physical mechanisms governing membrane motion---and their coupling to membrane geometry---remain poorly understood.

Lipid membranes are unique materials: lipids flow in-plane as a two-dimensional viscous fluid, yet the membrane bends out-of-plane as an elastic shell \cite{evans-skalak}.
Many theoretical and computational works neglect the in-plane shear viscosity when describing lipid membranes of various shapes and their stability \cite{ou-yang-pra-1989, seifert-pra-1991, fournier-prl-1996, seifert-long, steigmann-arma-1999, powers-pre-2002, capovilla-guven-jpa-2002, guven-jpa-2004, agrawal-steigmann-bmmb-2008, ramaswamy-prl-2014, al-izzi-prl-2018}, and consequently the in-plane viscous flow of lipids is often disregarded in the analysis of experimental results \cite{bar-ziv-prl-1994, goldstein-jp-1996, prost-prl-2002, kantsler-prl-2005, viallat-sm-2008, deschamps-prl-2009, keller-skalak-1982, kraus-prl-1996, seifert-epjb-1999, beaucourt-pre-2004, noguchi-prl-2007, vlahovska-pre-2007, messlinger-pre-2009, zhao-shaqfeh-jfm-2011, zhao-shaqfeh-pf-2011}.
However, the equations of motion governing arbitrarily curved and deforming lipid membranes, including all viscous and bending forces \cite{arroyo-pre-2009, powers-rmp-2010, kranthi-bmmb-2012, sahu-mandadapu-pre-2017} as well as additional irreversible phenomena \cite{sahu-mandadapu-pre-2017}, were recently obtained.
These equations show in-plane and out-of-plane membrane dynamics are nontrivially coupled through both continuity of the material and surface curvature.
Consequently, for example, the equations governing a flat sheet at the neuronal synapse are different from those describing spherical lipid membrane vesicles carrying chemical cargo in a shear flow, which are again different from the equations governing cylindrical tubes shooting from the endoplasmic reticulum.
Though the general equations of motion are known, there has not yet been a systematic effort to analyze which forces govern intramembrane flows and out-of-plane dynamics in various biologically relevant settings and geometries.

In this work, we study how lipid membrane geometry and dynamics are coupled.
We consider the three predominantly observed membrane geometries in biological systems: flat patches, spherical vesicles, and cylindrical tubes, each of which is a static solution to the equations governing lipid membrane dynamics.
For each geometry, the linearized dynamical lipid membrane equations are determined and non-dimensionalized via a scaling analysis.
In doing so, two dimensionless numbers are obtained.
The first is the familiar \FvK\ number ($\Gamma$), which compares tension and bending forces in the out-of-plane direction~\cite{nelson-pre-2003}.
The second is a new dimensionless number comparing viscous forces in the normal direction, which arise due to the coupling between in-plane viscous stresses and membrane curvature, to the well-known bending forces.
This quantity is named the Scriven--Love number ($\SL$) in honor of the seminal works on surface flows of arbitrarily curved two-dimensional fluids by L.E.~Scriven \cite{scriven-1960} and on elasticity of two-dimensional shells by A.E.H.~Love \cite{love}.
The Scriven--Love number is implicitly set to zero in studies which ignore viscous in-plane intramembrane flows.
In calculating $\SL$ in a variety of experimental studies, however, we find cases where viscous forces are non-negligible relative to bending forces in describing the dynamics of lipid membranes in response to shape perturbations.
Moreover, for spherical vesicles and cylindrical tubes we find experiments where
$\SL \gg 1$
and
$\Gamma \gg 1$,
such that bending forces contribute negligibly to the perturbed membrane's dynamical response.
In these situations, out-of-plane dynamics are governed by viscous and tension forces, which are highly coupled to in-plane flows, and the membrane behaves constitutively more like a soap bubble than an elastic shell.
Thus, it is necessary to consider the fluid nature of lipid membranes when understanding their behavior, shape, and dynamics in biological settings.

The remainder of this paper is organized as follows.
In Sec.~\ref{sec:sec_origin_scriven_love}, we review lipid membrane dynamics, describe how the in-plane membrane viscosity leads to a force in the out-of-plane direction, and present our calculation of the Scriven--Love number in several experimental studies.
Sections~\ref{sec:sec_flat_patch}, \ref{sec:sec_sphere}, and \ref{sec:sec_cylinder} describe the results from our analysis of flat patches, spheres, and cylinders, respectively.
We end with conclusions and avenues for future work in Sec.~\ref{sec:sec_conclusion}.
Detailed calculations of the unperturbed and perturbed equations for the different geometries, as well as the specifics of their non-dimensionalization, are provided in the Supplemental Material (SM)~\cite{supplemental}.

%
%

\section{Membrane Theory and Origin of the Scriven--Love Number} \label{sec:sec_origin_scriven_love}

The equations governing an arbitrarily curved and deforming lipid membrane are derived in the SM~\cite[Sec.~I]{supplemental}, following our previous irreversible thermodynamic developments~\cite{sahu-mandadapu-pre-2017}.
As lipid membranes can stretch only 2--3\% before tearing \cite{evans-skalak, nichol-jp-1996}, they are practically area incompressible, and are modeled as such.
The membrane is treated as a single differentiable manifold about the membrane mid-plane, implicitly assuming no slip between the two bilayer leaflets.
Inertial terms are not provided in the main text, as they are negligible in all geometries, for every case considered~\cite{supplemental}.
We also do not model the dynamics of the fluid surrounding the membrane, and include bulk effects only through the jump in the normal stress across the membrane surface.

The continuity equation of an area-incompressible lipid membrane is given by
\begin{equation} \label{eq:general_continuity}
	v^\alpha_{; \alpha}
	- 2 v H
	= 0
	~,
\end{equation}
where `$\alpha$' and other Greek indices span the set $\{ 1, 2 \}$ and denote independent directions on the surface.
In Eq.~\eqref{eq:general_continuity}, $v^\alpha$ are the two in-plane velocity components, $v$ is the normal velocity component, $H$ is the mean curvature, and $(\, \cdot \,)_{; \alpha}$ denotes the covariant derivative in the `$\alpha$' direction (details are provided in the SM~\cite[Sec.\ I.1]{supplemental}).
The continuity equation~\eqref{eq:general_continuity} indicates the surface divergence of the velocity field is zero, and this incompressibility constraint is enforced with the Lagrange multiplier $\lambda$---which physically acts as a surface tension, or equivalently the negative surface pressure.

The in-plane lipid membrane equations are found to be identical to those of a two-dimensional fluid film \cite{edwards-brenner}, and are given by
\begin{equation} \label{eq:general_in_plane_equations}
	0
	= \big(
		\lambda \, a^{\alpha \beta}
		+ \pi^{\beta \alpha}
	\big)_{; \beta}
	= a^{\alpha \beta} \lambda_{, \beta}
	+ \pi^{\beta \alpha}_{; \beta}
	~,
\end{equation}
where $a^{\alpha \beta}$ are the contravariant metric tensor components, $\lambda$ is the surface tension enforcing Eq.~\eqref{eq:general_continuity}, $\pi^{\alpha \beta}$ are the in-plane viscous stresses, and $(\lambda \, a^{\alpha \beta} + \pi^{\beta \alpha})$ are the total in-plane fluid stresses.
The notation $(\, \cdot \,)_{, \beta}$ denotes the partial derivative in the `$\beta$' direction.
Physically, the in-plane equations~\eqref{eq:general_in_plane_equations} indicate surface tension gradients balance the divergence of the in-plane viscous stresses, or equivalently the divergence of the in-plane fluid stresses are zero, analogous to the Stokes equations for a three-dimensional bulk fluid.

The out-of-plane equation governing lipid membrane dynamics, called the shape equation, is given by
\begin{equation} \label{eq:general_shape_equation}
	\hspace{4pt}
	0
	= p
	+ \big(
		\lambda \, a^{\alpha \beta}
		+ \pi^{\alpha \beta}
	\big) b_{\alpha \beta}
	- 2 \kb H \big(H^2 - K\big)
	- \kb \Delta_{\mathrm{s}} H
	~.
	\hspace{4pt}
\end{equation}
In Eq.~\eqref{eq:general_shape_equation}, $p$ is the jump in the normal stress across the membrane surface, $K$ is the Gaussian curvature, $\kb$ is the mean bending modulus, $\Delta_{\mathrm{s}}$ is the surface Laplacian operator, and $b_{\alpha \beta}$ are the curvature tensor components.
In the limit where there are no in-plane viscous stresses ($\pi^{\alpha \beta} = 0$, for example when the membrane is stationary) and no mean bending modulus $(\kb = 0)$, the shape equation \eqref{eq:general_shape_equation} reduces to the Young--Laplace equation
$p + \lambda \, a^{\alpha \beta} b_{\alpha \beta} = p + 2 \lambda H = 0$.
Next consider the bending terms in the shape equation~\eqref{eq:general_shape_equation}, which are expected as the membrane bends elastically out-of-plane.
In this case, bending terms arise from a free energy for two-dimensional shells in which there is no in-plane shear modulus to account for the fluidity of the lipid bilayer, as first put forth by P.B.\ Canham~\cite{canham-jtb-1970}, W.\ Helfrich~\cite{helfrich-1973}, and E.A.\ Evans~\cite{evans-bpj-1974}.
Finally, consider the $\pi^{\alpha \beta} b_{\alpha \beta}$ term in the shape equation \eqref{eq:general_shape_equation}, which is the contraction of the in-plane viscous stresses with the membrane curvature and was also found in the study of fluid film dynamics \cite{edwards-brenner}.
It is this term which, when compared to the bending forces, gives rise to the Scriven--Love number; we discuss its origin and physical consequences below.

\begin{figure}[t!]
	\centering
	\subfigure[\ tension forces: $\lambda \, a^{\alpha \beta} b_{\alpha \beta} = 2 \lambda H$]{\includegraphics[width=0.78\linewidth]{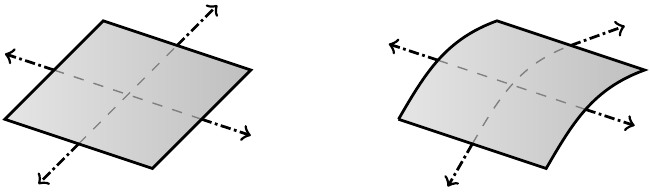}
	\label{fig:fig_scriven_love_schematic_tension}}

	\subfigure[\ viscous forces: $\pi^{\alpha \beta} b_{\alpha \beta}$]{\includegraphics[width=0.78\linewidth]{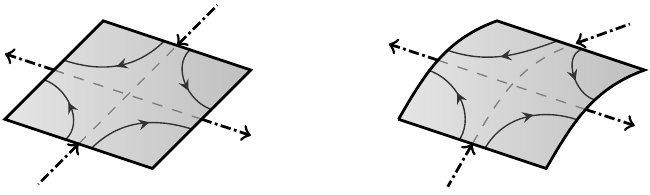}
	\label{fig:fig_scriven_love_schematic_viscous}}
	\caption{
		A schematic showing how surface tension and viscous forces arise in the normal direction, with dashed arrows depicting boundary tractions.
		(a) The surface tension $\lambda$ pulls the membrane at each edge, such that when the shape is perturbed a normal force $2 \lambda H$ arises.
		(b) In an extensional flow, the velocity field is given by
		$\bmv = \dot{\gamma} ( x \bm{e}_x - y \bm{e}_y)$,
		as depicted by the solid arrows.
		The tractions push and pull the fluid, as shown with the dashed arrows, such that when the membrane is perturbed a viscous force $\pi^{\alpha \beta} b_{\alpha \beta}$ arises in the normal direction.
	}
	\label{fig:fig_scriven_love_schematic}
\end{figure}

To understand how in-plane viscous stresses lead to out-of-plane forces, we first recognize the $2 \lambda H$ term in the shape equation \eqref{eq:general_shape_equation}, often called the Laplace pressure, arises due to the surface tension acting in different directions at different locations---despite the stress associated with the surface tension, $\lambda \, a^{\alpha \beta}$, being in-plane and isotropic (see Fig.\ \ref{fig:fig_scriven_love_schematic_tension}).
We next consider the shear stresses arising from a planar extensional flow, for which streamlines and boundary tractions are shown on the left-hand side of Fig.~\ref{fig:fig_scriven_love_schematic_viscous} (solid and dotted arrows, respectively).
Just as in the case of the surface tension, when the membrane is curved the viscous stresses act in different directions at different locations on the surface and give rise to the $\pi^{\alpha \beta} b_{\alpha \beta}$ force in the normal direction (Fig.~\ref{fig:fig_scriven_love_schematic_viscous}).
Our analysis of this scenario thus leads to a general conclusion: for a curved surface, in-plane stresses lead to out-of-plane forces.
Following similar arguments, one can show out-of-plane shearing tractions couple to membrane curvature to produce a resulting in-plane force.
A detailed description of the nontrivial in-plane and out-of-plane coupling of lipid membrane forces, from the perspective of balance laws and irreversible thermodynamics, is presented in the SM \cite[Secs.\ I.2, I.3]{supplemental}.

The membrane shape equation~\eqref{eq:general_shape_equation} has contributions from four different forces, arising from the jump in the bulk normal stress, surface tension, mean bending modulus, and in-plane viscosity.
In our analysis, we generally assume a base state with no viscous stresses, such that the normal stress jump, surface tension, and bending forces balance.
When the membrane shape is perturbed, viscous forces result in the normal direction and a natural question arises: How much do viscous forces contribute to the membrane's dynamical response?
This question is addressed via a scaling analysis of the unperturbed and perturbed equations.

\def\arraystretch{1.3}
\setlength\tabcolsep{7pt}
\begin{table*}[t!]
	\begin{center}
		\def~{\hphantom{0}}
		\begin{tabular}{c | c c c c c c c c}
			\hline
			\hline
			~ &
			\multirow{2}{*}{Ref.} &
			\multirow{2}{*}{$\SL$} &
			\multirow{2}{*}{$\Gamma$} &
			\multirow{2}{*}{Symbol} &
			\multirow{2}{*}{$V$ (nm/$\mu$s)} &
			\multirow{2}{*}{$R$ (nm)} &
			\multirow{2}{*}{$\kb$ (pN$\cdot$nm)} &
			\multirow{2}{*}{$\Lambda$ (pN/nm)}
			\\[-1pt]
			~ &
			&
			&
			&
			&
			&
			&
			&
			~
			\\[1pt]
			\hline
			\multirow{3}{*}{\includegraphics[width=0.1\linewidth]{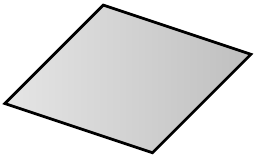}} &
			\cite{watanabe-elife-2013} &
			$8 \cdot 10^{-3}$ &
			\, $1 \cdot 10^{-1}$ &
			$\blacktriangle$ &
			$8 \cdot 10^{-4}$ &
			$1 \cdot 10^2$ &
			-- &
			--
			\\
			&
			\cite{watanabe-elife-2013} &
			$2 \cdot 10^{-4}$ &
			\, $2 \cdot 10^{-2}$ &
			$\bigstar$ &
			$3 \cdot 10^{-5}$ &
			$5 \cdot 10^1$ &
			-- &
			--
			\\
			&
			\cite{cocucci-cell-2012} &
			$6 \cdot 10^{-4}$ &
			$1 \cdot 10^{1}$ &
			$\blacksquare$ &
			$6 \cdot 10^{-6}$ &
			$1 \cdot 10^3$ &
			-- &
			--
			\\[1pt]
			\hline
			\multirow{5}{*}{\includegraphics[width=0.08\linewidth]{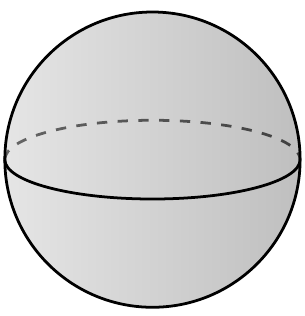}} &
			\cite{mellema-pre-1997} &
			$1 \cdot 10^4$ &
			$5 \cdot 10^5$ &
			$\bowtie$ &
			$6 \cdot 10^{-1}$ &
			$6 \cdot 10^4$ &
			30 &
			$4 \cdot 10^{-3}$
			\\
			&
			\cite{lipowsky-mr-1980, kolaczkowska-nri-2013} &
			$2 \cdot 10^3$ &
			$4 \cdot 10^2$ &
			$\clubsuit$ &
			$3$ &
			$6 \cdot 10^3$ &
			-- &
			--
			\\
			&
			\cite{ota-ac-2009} &
			$4 \cdot 10^2$ &
			$4 \cdot 10^2$ &
			$\diamondsuit$ &
			$6 \cdot 10^{-1}$ &
			$6 \cdot 10^3$ &
			-- &
			--
			\\
			&
			\cite{mader-epje-2006} &
			$2 \cdot 10^1$ &
			$6 \cdot 10^2$ &
			$\heartsuit$ &
			$3 \cdot 10^{-2}$ &
			$1 \cdot 10^4$ &
			170 &
			--
			\\
			&
			\cite{lipowsky-mr-1980, forster-pnas-2005} &
			\, $1 \cdot 10^{-1}$ &
			\, $2 \cdot 10^{-2}$ &
			$\spadesuit$ &
			$3 \cdot 10^{-3}$ &
			$5 \cdot 10^1$ &
			-- &
			--
			\\[1pt]
			\hline
			\multirow{4}{*}{\includegraphics[width=0.05\linewidth]{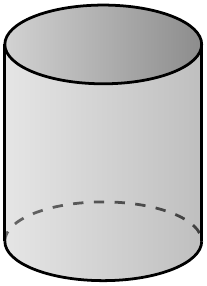}} &
			\cite{shi-cell-2018} &
			$7$ &
			$7$ &
			$\maltese$ &
			$2$ &
			$1 \cdot 10^2$ &
			$380$ &
			$2 \cdot 10^{-1}$
			\\
			&
			\cite{shi-cell-2018} &
			$1$ &
			$1$ &
			$\circledast$ &
			$1$ &
			$4 \cdot 10^1$ &
			$380$ &
			$3 \cdot 10^{-1}$
			\\
			&
			\cite{shi-cell-2018} &
			$\bfrac{1}{4}$ &
			$\bfrac{1}{4}$ &
			$\P$ &
			$4 \cdot 10^{-1}$ &
			$2 \cdot 10^1$ &
			$250$ &
			$2 \cdot 10^{-1}$
			\\
			&
			\cite{shi-cell-2018} &
			\, $4 \cdot 10^{-3}$ &
			$2$ &
			$\twonotes$ &
			$4 \cdot 10^{-3}$ &
			$4 \cdot 10^1$ &
			$380$ &
			$6 \cdot 10^{-1}$
			\\
			\hline
			\hline
		\end{tabular}
		\caption{Calculation of the Scriven--Love number $\SL$ and \FvK\ number $\Gamma$ in several experimental studies involving flat patches, spherical vesicles, and cylindrical tubes.
		Values of $\SL$ and $\Gamma$ are calculated according to Eqs.\ \eqref{eq:general_scriven_love} and \eqref{eq:foppl_von_karman_def} in planar geometries, Eqs.\ \eqref{eq:foppl_von_karman_sphere} and \eqref{eq:sphere_scriven_love} in spherical geometries, and Eq.\ \eqref{eq:cylinder_scriven_love} in cylindrical geometries.
		In the first three cylindrical experiments ($\maltese$, $\circledast$, $\P$), the base velocity is zero and the perturbed velocity scale is reported according to Eq.\ \eqref{eq:cylinder_initially_static_general_scalings}$_3$.
		The column titled `$R$' denotes the radius in spherical and cylindrical geometries, and the characteristic length scale in planar geometries.
		The same data are shown visually in Fig.\ \ref{fig:fig_plot}, using the same symbols.
		When unreported, the characteristic values $\kb = 100$ pN$\cdot$nm and $\Lambda = 10^{-3}$ pN/nm were used; in all cases membrane viscosities were not explicitly measured and a characteristic value
		$\zeta = 10$ pN$\cdot \mu$sec/nm \cite{honerkamp-prl-2013} was used (see Table \ref{tab:tab_parameters}).
		Our detailed calculation of characteristic values in each experimental study is provided in the SM \cite{supplemental}.}
		\vspace{-13pt}
		\label{tab:tab_experiments}
	\end{center}
\end{table*}

A scaling analysis of the continuity \eqref{eq:general_continuity}, in-plane \eqref{eq:general_in_plane_equations}, and shape \eqref{eq:general_shape_equation} equations in various geometries reveals the competition between viscous and bending forces in the perturbed equations leads to the Scriven--Love number $\SL$, which is of the general form
\begin{equation} \label{eq:general_scriven_love}
	\SL
	\, = \, \dfrac{O(\pi^{\alpha \beta} b_{\alpha \beta})}{O(\kb \Deltas H)}
	\, = \, \dfrac{\zeta V L}{\kb}
	~.
\end{equation}
In Eq.\ \eqref{eq:general_scriven_love},
$\zeta$ is the coefficient of in-plane membrane viscosity,
$V$ is a characteristic velocity scale,
and
$L$ is the size of the system: either the length of a flat patch, or the radius of a spherical vesicle or cylindrical tube.
Comparing bending and surface tension forces in the normal direction also leads to the \FvK\ number $\Gamma$~\cite{nelson-pre-2003}, given by
\begin{equation} \label{eq:foppl_von_karman_def}
	\Gamma
	= \dfrac{\Lambda L^2}{\kb}
	~,
\end{equation}
where $\Lambda$ is a characteristic surface tension scale.
The results of our non-dimensionalization for planar, spherical, and cylindrical geometries, including the form of the Scriven--Love and \FvK\ numbers, are provided in Secs.~\ref{sec:sec_flat_patch}--\ref{sec:sec_cylinder}; the details of our scaling analysis are again provided in the SM \cite{supplemental}.
We note that the \FvK\ number was found when characterizing the membrane equations in previous studies, in which it is also referred to as a dimensionless tension~\cite{boedec-jfm-2014, narsimhan-shaqfeh-jfm-2015}.

\begin{figure}[b!]
	\vspace{-32pt}
	\centering
	\includegraphics[width=0.98\linewidth]{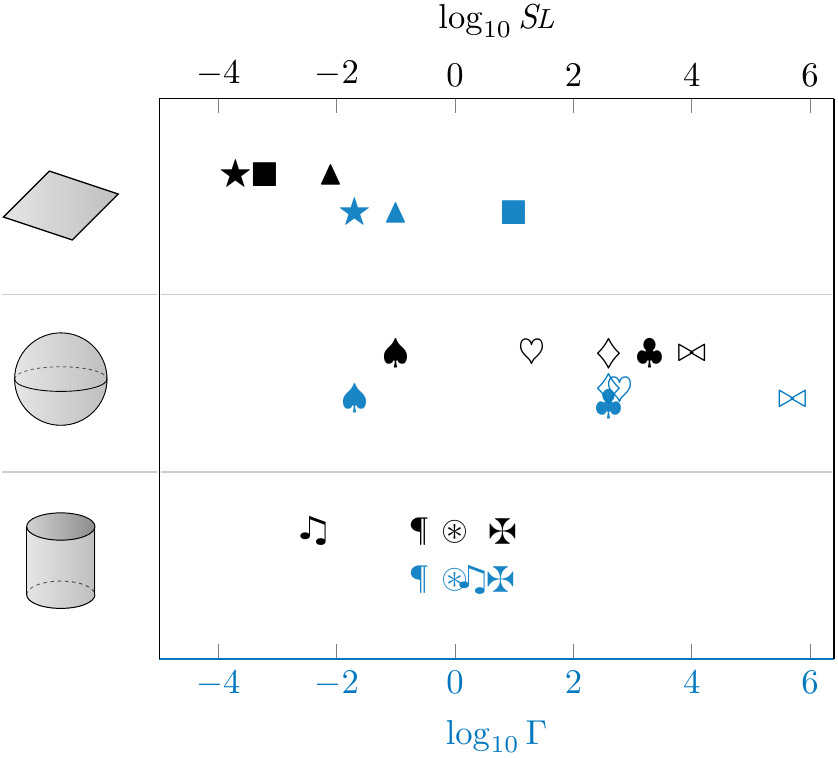}
	\vspace{-3pt}
	\caption{
		Plot of the Scriven--Love and \FvK\ numbers in various experimental studies.
		For each experimental result in Table \ref{tab:tab_experiments}, the Scriven--Love number (black, above) and \FvK\ number (blue, below) is plotted using the same symbol.
		In planar membranes, we only found experiments where
		$\SL \ll 1$,
		while $\Gamma$ ranged from $10^{-2}$ to 10.
		Spherical vesicles were found to have a wide range of both dimensionless numbers, such that the dynamics of a perturbed vesicle can differ significantly between different experiments.
		For cylindrical tubes, tension forces were always significant, and in all but one case the Scriven--Love number was non-negligible as well.
		Additional details for each geometry are provided in corresponding sections in the main text.
	}
	\vspace{-5pt}
	\label{fig:fig_plot}
\end{figure}

Our calculation of the Scriven--Love and \FvK\ numbers for many experimental studies involving different lipid membrane geometries is provided in Table~\ref{tab:tab_experiments}; the same data are shown visually in Fig.~\ref{fig:fig_plot}.
While there are a wealth of experimental studies of membrane behaviors in different geometries, many studies do not report characteristic velocity scales, and so we are unable to quantify the Scriven--Love number in those cases.
Moreover, the calculation of membrane tension in many experimental studies of lipid membrane tubes fails to account for the possibility of a pressure drop across the membrane surface.
Accordingly, in many cases the collected data are insufficient to quantify the membrane tension---which, as a Lagrange multiplier enforcing areal incompressibility, takes the requisite value at every point on the membrane surface to locally satisfy the incompressibility constraint (see discussion in Sec.~\ref{sec:sec_cylinder}).
In several of our calculations in Table~\ref{tab:tab_experiments}, bending moduli and tension values were not reported; in those cases we used the characteristic values provided in Table~\ref{tab:tab_parameters}.
We hope our finding of large $\SL$ and $\Gamma$ in many experimental works motivates the simultaneous reporting of characteristic velocity and tension scales in future studies.

In the following sections, we provide and analyze the membrane equations in planar, spherical, and cylindrical geometries.
These geometries are
(i) commonly found in biological settings,
(ii) relevant to many in vitro studies of lipid bilayers, and
(iii) static solutions to the membrane equations \eqref{eq:general_continuity}--\eqref{eq:general_shape_equation}.
In situations where experimental data are available, we refer to experimental studies according to the symbols provided in Table~\ref{tab:tab_experiments} and Fig.~\ref{fig:fig_plot}.

%
%

\section{Flat Membrane Patches} \label{sec:sec_flat_patch}

\begin{figure}[!b]
	\vspace{-10pt}
	\centering
	\subfigure[\ unperturbed: $\bmx_\sz (x, y)$]{\includegraphics[width=0.45\linewidth]{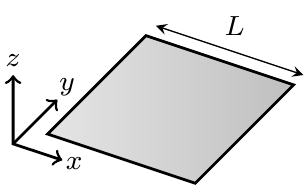}
	\label{fig:fig_flat_unperturbed}}
	\hspace{10pt}
	\subfigure[\ perturbed: $\bmx (x, y, t)$]{\includegraphics[width=0.38\linewidth]{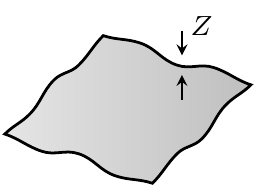}
	\label{fig:fig_flat_perturbed}}
	\caption{
		Schematic of the unperturbed (a) and perturbed (b) flat plane geometries.
		The membrane patch has a characteristic length $L$, and perturbations are of a characteristic height $Z$, with $\epsilon := Z / L \ll 1$.
	}
	\label{fig:fig_flat}
\end{figure}

In this section, we consider nearly planar lipid membranes, with either
(i) no base flow or
(ii) an in-plane base flow.
In both cases, the initial membrane position is given by
\begin{equation} \label{eq:flat_unperturbed_position}
	\bm{x}_\sz (x, y) 
	= x \, \bm{e}_x
	+ y \, \bm{e}_y
	~,
\end{equation}		
where $x$ and $y$ are standard Cartesian coordinates, and the patch size sets the length scale $L$ (see Fig.~\ref{fig:fig_flat_unperturbed}).
In Eq.~\eqref{eq:flat_unperturbed_position} and from now on, a subscript or superscript `$\sz$' denotes an unperturbed quantity.
We denote the unperturbed velocity components and surface tension in the base state as $v^x_\sz$, $v^y_\sz$, $v_\sz$, and $\lambda_\sz$.
As the membrane base state shape is fixed, the unperturbed normal velocity
$v_\sz = 0$, as will be the case for the unperturbed spherical and cylindrical geometries as well.

We next introduce a height perturbation in the normal direction, such that the perturbed membrane position is given by
\begin{equation} \label{eq:flat_perturbed_position}
	\bm{x}(x, y, t) 
	\, = \, x \, \bm{e}_x
	\, + \, y \, \bm{e}_y
	\, + \, \epsilon \, \tilh (x,y,t) \, \bm{e}_z
	~.
\end{equation}
In Eq.~\eqref{eq:flat_perturbed_position}, the total height perturbation $\epsilon \, \tilh (x, y, t)$ is assumed to be $O(Z)$, where
$Z \ll L$.
In this case,
$\epsilon := Z / L$
is a small parameter and $\tilh$ is $O(L)$, as depicted in Fig.~\ref{fig:fig_flat_perturbed}.
The surface parametrization in Eq.\ \eqref{eq:flat_perturbed_position} is commonly used in the study of nearly planar surfaces, and is called the Monge parametrization \cite{monge}.
The membrane velocity components and surface tension are expanded as
\begin{equation} \label{eq:flat_unknown_expansions}
	\begin{split}
		v^x
		&= v^x_\sz
		+ \epsilon \, \tilv^x
		~,
		\qquad \qquad \hspace{6pt}
		v^y
		= v^y_\sz
		+ \epsilon \, \tilv^y
		~,
		\\[8pt]
		v
		&= \epsilon \, \tilh_{, t}
		~,
		\qquad \quad
		\text{and}
		\qquad \quad
		\lambda
		= \lambda_\sz
		+ \epsilon \, \till
		~,
	\end{split}
\end{equation}
where the parameter $\epsilon$ captures the smallness of the perturbations to the velocity components and surface tension.

%
%

\subsection{Initially Static Membrane Patch} \label{sec:sec_flat_initially_static_membrane}

For a flat, initially static membrane,
$v^x_\sz = 0$,
$v^y_\sz = 0$,
$v_\sz = 0$,
and
$\lambda_\sz = \lambda^{}_0$,
where $\lambda^{}_0$ is a constant set by the base state boundary conditions.
In this case, the unperturbed membrane state sets the surface tension scale $\Lambda$ as
\begin{equation} \label{eq:flat_static_unperturbed_surface_tension_scale}
	\Lambda
	:= \lambda^{}_0
	~,
\end{equation}
and also sets the length scale $L$ over which quantities vary.
The in-plane velocity scale $V$ is determined via a simple scaling analysis after the perturbed equations are non-dimensionalized.

The derivation of the linearized perturbed governing equations, and their subsequent non-dimensionalization, is algebraically involved and hence relegated to the SM \cite[Secs.\ II.2, II.3]{supplemental}; see also Ref.\ \cite{seifert-epl-1993}.
We non-dimensionalize equations with the dimensionless quantities introduced in Table~\ref{tab:tab_flat_dimensionless_definitions}.
The dimensionless perturbed continuity, in-plane $x$, in-plane $y$, and shape equations, to first order in $\epsilon$, are respectively given by
\begin{gather}
	\tilv^{x \nd}_{, \xnd}
	\, + \, \tilv^{y \nd}_{, \ynd}
	\, = \, 0
	~,
	\label{eq:flat_static_perturbed_nd_continuity}
	\\[6pt]
	\tilv^{x \nd}_{, \xnd \! \xnd}
	\, + \, \tilv^{x \nd}_{, \ynd \! \ynd}
	\, + \, \tillnd_{, \xnd}
	\, = \, 0
	~,
	\label{eq:flat_static_perturbed_nd_x}
	\\[6pt]
	\tilv^{y \nd}_{, \xnd \! \xnd}
	\, + \, \tilv^{y \nd}_{, \ynd \! \ynd}
	\, + \, \tillnd_{, \ynd}
	\, = \, 0
	~,
	\label{eq:flat_static_perturbed_nd_y}
	\\[-6pt]
	\intertext{and}
	\Gamma \, \Deltand \, \tilhnd
	\, - \, \dfrac{1}{2} \, (\Deltand)^2 \, \tilhnd
	\, = \, 0
	~.
	\label{eq:flat_static_perturbed_nd_shape}
\end{gather}

\def\arraystretch{1.3}
\setlength\tabcolsep{4pt}
\begin{table}[t!]
	\begin{center}
		\def~{\hphantom{0}}
		\begin{tabular}{l c c r}
			\hline
			\hline
			Parameter &
			Symbol &
			Value &
			Ref.
			\\[1pt]
			\hline
			intramembrane viscosity &
			$\zeta$ &
			10 pN$\cdot \mu$s/nm &
			\cite{honerkamp-prl-2013}
			\\
			mean bending modulus &
			$\kb$ &
			100 pN$\cdot$nm &
			\cite{pecreaux-bassereau-epje-2004}
			\\
			low surface tension &
			$\Lambda$ &
			$10^{-4}$ pN/nm &
			\cite{pecreaux-bassereau-epje-2004}
			\\
			high surface tension &
			$\Lambda$ &
			$10^{-1}$ pN/nm &
			\cite{dai-jn-1998}
			\\
			\hline
			\hline
		\end{tabular}
		\caption{
			Characteristic membrane material parameters.
			The surface tension can span a wide range of values to prevent significant areal dilation; characteristic high and low values are provided.
		}
		\vspace{-14pt}
		\label{tab:tab_parameters}
	\end{center}
\end{table}

\noindent From the in-plane equations (\ref{eq:flat_static_perturbed_nd_x}, \ref{eq:flat_static_perturbed_nd_y}) the velocity scale $V$ is found to be
\begin{equation} \label{eq:flat_static_velocity_time_scaling}
	V
	\, = \, \dfrac{L \Lambda}{\zeta}
	~,
\end{equation}
such that in-plane viscous forces balance tension gradients.
In Eq.\ \eqref{eq:flat_static_perturbed_nd_shape}, the \FvK\ number $\Gamma$ is defined as in Eq.~\eqref{eq:foppl_von_karman_def} and the surface Laplacian of a scalar quantity is given by
$\Delta_{\mathrm{s}} (\, \cdot \,) = (\, \cdot \,)_{, x x} + (\, \cdot \,)_{, y y}$.
The continuity~\eqref{eq:flat_static_perturbed_nd_continuity} and in-plane~(\ref{eq:flat_static_perturbed_nd_x}, \ref{eq:flat_static_perturbed_nd_y}) equations are familiar from the study of incompressible, low-Reynolds number bulk fluids.
The shape equation \eqref{eq:flat_static_perturbed_nd_shape} contains two terms: the first is the out-of-plane surface tension force $2 \lambda H$, and the second is the bending force $- \kb \Delta_{\mathrm{s}} H$ (cf.~Eq.~\eqref{eq:general_shape_equation}).
Note that as the unperturbed membrane has no in-plane viscous stresses
($\pi^{\alpha \beta}_\sz = 0$)
and no curvature
($b_{\alpha \beta}^\sz = 0$),
the Scriven--Love number does not appear in the shape equation \eqref{eq:flat_static_perturbed_nd_shape}.
Thus, viscous forces in the normal direction are irrelevant in a perturbed, initially static planar membrane.

\def\arraystretch{2.0}
\setlength\tabcolsep{10pt}
\begin{table}[t!]
	\begin{center}
		\def~{\hphantom{0}}
		\begin{tabular}{|c|c|c|c|}
			\hline
			\hline
			$
				x^\nd
				:= \dfrac{x}{L}
			$
			&
			$
				y^\nd
				:=
				\dfrac{y}{L}
			$
			&
			$
				\tilh^\nd
				:=
				\dfrac{\tilh}{L}
				\rule{0mm}{1.9em}
			$
			\\[7pt]
			\hline
			$
				\tilv^{x \nd}
				:= \dfrac{\tilv^x}{V}
			$
			&
			$
				\tilv^{y \nd}
				:= \dfrac{\tilv^y}{V}
			$
			&
			$
				\till^*
				:= \dfrac{\till}{\Lambda}
				\rule{0mm}{1.9em}
			$
			\\[7pt]
			\hline
			\hline
		\end{tabular}
		\caption{
			Dimensionless definitions for an initially planar lipid membrane.
			Here, $L$ is the length of the membrane patch and $V$ is a characteristic velocity scale.
		}
		\vspace{-5pt}
		\label{tab:tab_flat_dimensionless_definitions}
	\end{center}
\end{table}

For lipid and biological membranes,
$\kb \sim 100$ pN$\cdot$nm \cite{pecreaux-bassereau-epje-2004}
and we consider membrane patches of side length
$L \sim 100$--1000 nm,
as relevant for the experiments in Table~\ref{tab:tab_experiments}.
The surface tension scale $\Lambda$ is set in the base state, and can be arbitrarily small or large as required by the membrane's areal incompressibility constraint---up to values as large as $\sim$1--10 pN/nm, at which point the membrane tears~\cite{evans-skalak, nichol-jp-1996}.
For example, the base tension in neurons is estimated to be $\sim \! 10^{-2}$ pN/nm in isotonic conditions, yet tensions of $\sim \! 10^{-1}$ pN/nm are observed in neurons placed in hypotonic solutions~\cite{dai-jn-1998}.
Moreover, surface tensions as low as $\sim \! 10^{-4}$ pN/nm were observed in giant unilamellar vesicles (GUVs)~\cite{pecreaux-bassereau-epje-2004}, and thus we expect
$\Lambda \sim 10^{-4}$--$10^{-1}$ pN/nm
in different flat lipid membrane patches.
At low tensions of
$\Lambda \sim 10^{-4}$ pN/nm
with
$L \sim 100$ nm,
$\Gamma \sim 10^{-2} \ll 1$
and bending dominates the shape equation~\eqref{eq:flat_static_perturbed_nd_shape}---which simplifies to
$(\Deltand)^2 \, \tilhnd  = 0$,
such that there is no coupling between in-plane and out-of-plane dynamics.
At high tensions of
$\Lambda \sim 10^{-1}$ pN/nm
with
$L \sim 10^3$ nm,
$\Gamma \sim 10^3 \gg 1$
and the shape equation
$\Deltand \, \tilhnd = 0$
is tension-dominated as for a fluid film or soap bubble.
Finally, at moderate tensions of
$\Lambda \sim 10^{-2}$ pN/nm
and length scales $L \sim 100$ nm,
$\Gamma \sim 1$ and the surface tension and bending terms balance in governing the membrane shape.
In the latter two cases, the surface tension provides the only coupling between in-plane and out-of-plane dynamics, and in all three cases viscous forces play no role in determining the membrane shape.

%
%

\subsection{Membrane Patch with a Base Flow} \label{sec:sec_flat_base_flow}

We now analyze an initially planar membrane with a simple Couette base flow:
$\bmv_\sz = y \, v^{}_0 / L \, \bm{e}_x$
and
$\lambda_\sz = \lambda^{}_0$,
such that
$v^x_{\sz, y} = v^{}_0 / L$.
As in the initially static case, the base state boundary conditions set the surface tension scale
$\Lambda = \lambda^{}_0$,
which is independent of $V$ and $L$.
However, the base flow sets the characteristic velocity scale
$V = v^{}_0$,
and we can no longer choose the velocity scale such that perturbed in-plane viscous and tension forces balance.
In what follows, we use a scaling analysis to demonstrate the existence of a new length scale $\ell$ over which $\tilv^x$, $\tilv^y$, and $\till$ vary such that in-plane forces balance; moreover, $\tilh$ is found to vary not over the new length scale $\ell$ but rather over the patch size $L$.
While we consider only this particular base flow in the main text, the scaling analysis of the perturbed equations given a general base flow is provided in the SM \cite[Sec.\ II.4]{supplemental}.

Suppose the perturbed in-plane velocities and surface tension vary over the patch size $L$, as was assumed for an initially static membrane in Sec.~\ref{sec:sec_flat_initially_static_membrane}.
As shown in the SM \cite[Sec.\ II.4.b]{supplemental}, non-dimensionalization of the perturbed in-plane equations yields
$
	(\Lambda L / (\zeta V)) \tillnd_{, \xnd}
	+ \, \tilv^{x \nd}_{, \xnd \! \xnd}
	+ \, \tilv^{x \nd}_{, \ynd \! \ynd}
	= \, 0
$
and
$
	(\Lambda L / (\zeta V)) \tillnd_{, \ynd}
	+ \, \tilv^{y \nd}_{, \xnd \! \xnd}
	+ \, \tilv^{y \nd}_{, \ynd \! \ynd}
	= \, 0
$.
In the limit where
$V \ll \Lambda L / \zeta$,
the in-plane equations imply
$\tillnd = \text{constant}$,
such that surface tension gradients no longer balance in-plane viscous forces.
However, in the limit where the base velocity $V$ tends to zero, we expect to recover Eqs.\ \eqref{eq:flat_static_perturbed_nd_x} and \eqref{eq:flat_static_perturbed_nd_y}, namely, the perturbed in-plane equations for an initially static patch.
As these equations are not recovered in this case, the solution
$\tillnd = \text{constant}$
in the limit of small $V$ is unphysical, and our assumption that all quantities vary over a length scale $L$ is incorrect.

We next assume there exists some new length scale $\ell$ over which perturbed quantities vary.
In this case, in-plane viscous forces are
$O(\zeta \tilv^x_{, x x}) = \zeta V / \ell^2$
and in-plane surface tension forces are
$O(\till_{, x}) = \Lambda / \ell$; equating the two such that viscous and tension forces are of the same order reveals a new characteristic length
\begin{equation} \label{eq:flat_base_flow_perturbed_length_scale}
	\ell
	\, = \, \dfrac{\zeta V}{\Lambda}
	~.
\end{equation}
Assuming out-of-plane height perturbations vary over the length scale $\ell$, such that
$O(\Deltas \tilh) = L / \ell^2$,
yields the shape equation
$
	(\zeta^2 V^2 / (\kb \Lambda))
	(2 \tilhnd_{, \xnd \! \ynd}
	+ \Deltand \tilhnd)
	- \tfrac{1}{2} \Deltandd \tilhnd
	= 0
$.
However, in the limit of vanishing base state velocity, the shape equation simplifies to
$ \Deltandd \tilhnd = 0 $,
and the initially static result \eqref{eq:flat_static_perturbed_nd_shape} is again not recovered.
Consequently, our assumption that all quantities vary over the new length scale $\ell$ is also incorrect.

At this point, though neither of our scaling attempts thus far were valid, we realize
(i) the problem requires a new length scale over which $\tilv^x$, $\tilv^y$, and $\till$ vary, and
(ii) the out-of-plane perturbed shape $\tilh$ cannot also vary over the same length scale.
We therefore posit that while $\tilv^x$, $\tilv^y$, and $\till$ vary over the length scale $\ell$ \eqref{eq:flat_base_flow_perturbed_length_scale},
$\tilh$ varies over the patch length $L$.
Due to the presence of two different lengths in the scaling analysis, we define the new dimensionless quantities
\begin{equation} \label{eq:flat_base_flow_new_length_scaling}
	\xndl
	:= \, \dfrac{x}{\ell}
	\qquad
	\text{and}
	\qquad
	\yndl
	:= \, \dfrac{y}{\ell}
	~.
\end{equation}
As shown in Sec.\ II.4 of the SM \cite{supplemental}, with this scaling result the dimensionless perturbed first-order governing equations are found to be
\begin{gather} 
	\tilv^{x \nd}_{, \xndl}
	\, + \, \tilv^{y \nd}_{, \yndl}
	= \, 0
	~,
	\label{eq:flat_dynamic_perturbed_continuity_nd}
	\\[8pt]
	\tilv^{x \nd}_{, \xndl \! \xndl}
	\, + \, \tilv^{x \nd}_{, \yndl \! \yndl}
	\, + \, \tillnd_{, \xndl}
	\, = \, 0
	~,
	\label{eq:flat_dynamic_perturbed_in_plane_x_nd}
	\\[8pt]
	\tilv^{y \nd}_{, \xndl \! \xndl}
	\, + \, \tilv^{y \nd}_{, \yndl \! \yndl}
	\, + \, \tillnd_{, \yndl}
	\, = \, 0
	~,
	\label{eq:flat_dynamic_perturbed_in_plane_y_nd}
	\\[-8pt]
	\intertext{and}
	2 \SL \,\, \tilhnd_{, \xnd \! \ynd}
	\, + \, \Gamma \, \Deltand \, \tilhnd
	\, - \, \dfrac{1}{2} \, (\Deltand)^2 \, \tilhnd
	\, = \, 0
	~.
	\label{eq:flat_dynamic_perturbed_shape_nd}
\end{gather}
In Eq.\ \eqref{eq:flat_dynamic_perturbed_shape_nd}, the Scriven--Love number
$\SL = \zeta V L / \kb$ \eqref{eq:general_scriven_love}
and the \FvK\ number
$\Gamma = \Lambda L^2 / \kb$ \eqref{eq:foppl_von_karman_def}.

In comparing Eqs.~\eqref{eq:flat_dynamic_perturbed_continuity_nd}--\eqref{eq:flat_dynamic_perturbed_shape_nd} to their counterparts in the initially static case~\eqref{eq:flat_static_perturbed_nd_continuity}--\eqref{eq:flat_static_perturbed_nd_shape}, we make several observations.
First, the continuity and in-plane equations now involve spatial derivatives over the length scale $\ell$, rather than the patch length $L$.
As a result, in-plane viscous and tension forces are both $O(\Lambda^2 / (\zeta V))$, a scaling which is difficult to predict from a simple non-dimensionalization of the governing equations.
Second, the length scale $\ell$ satisfies the relation
$\ell = L \, \Gamma^{-1} \SL$,
such that the relative distance over which perturbed in-plane and out-of-plane quantities vary is set by the ratio of the Scriven--Love and \FvK\ numbers.
In the limit of $V$ going to zero, both $\SL$ and $\ell$ tend to zero and there is no longer a new length scale over which perturbed in-plane quantities vary.
In this case, the shape equation \eqref{eq:flat_dynamic_perturbed_shape_nd} simplifies to its initially static analog \eqref{eq:flat_static_perturbed_nd_shape}.
The Scriven--Love number emerges in Eq.~\eqref{eq:flat_dynamic_perturbed_shape_nd} due to the perturbed $\pi^{\alpha \beta} b_{\alpha \beta}$ term, which contains the coupling between in-plane shear stresses and membrane curvature.
Linearizing $\pi^{\alpha \beta} b_{\alpha \beta}$ in planar geometries yields
$\pi^{\alpha \beta}_\sz \, \tilh_{, \alpha \beta}$,
which in this case consists of only $\pi^{x y}_\sz \, \tilh_{, x y}$ due to our choice of the base velocity
$\bmv_\sz = y \, v^{}_0 / L \, \bm{e}_x$.
For a general base flow, all second derivatives of $\tilh$ are involved (see SM \cite[Sec.\ II.4]{supplemental}).
As shown in Table~\ref{tab:tab_experiments} and Fig.~\ref{fig:fig_plot},
$\SL \ll 1$
in all planar experimental systems considered, in which case the shape equation~\eqref{eq:flat_dynamic_perturbed_shape_nd} simplifies to that of an initially static membrane~\eqref{eq:flat_static_perturbed_nd_shape}.
As before, the dynamics of the perturbed membrane can be bending dominated \cite[$\blacktriangle$, $\bigstar$]{watanabe-elife-2013} or tension dominated \cite[$\blacksquare$]{cocucci-cell-2012}, or the tension and bending forces can balance.

At this point, we conclude our calculations for an initially flat membrane patch.
We note the \FvK\ number enters the shape equation both when the membrane is initially static \eqref{eq:flat_static_perturbed_nd_shape} and has a base flow~\eqref{eq:flat_dynamic_perturbed_shape_nd}; the Scriven--Love number appears only in the latter case.
Though we find viscous forces in the out-of-plane direction to be negligible in all experimental systems considered \cite{watanabe-elife-2013, cocucci-cell-2012}, the tension forces can be significant---and in some instances dominate bending forces in governing the perturbed membrane's dynamical response.

%
%

\section{Spherical Membrane Vesicles} \label{sec:sec_sphere}

We next consider spherical lipid membrane vesicles, which are found throughout the cell: vesicles are involved in endocytosis~\cite{mcmahon-cell-2002} and exocytosis~\cite{zhang-bpj-2010} as material is transported across the cell membrane, lysosomes fuse with food vacuoles to break down chemical compounds during phagocytosis~\cite{allen-coi-1996}, and transport vesicles shuttle proteins and lipids between the endoplasmic reticulum and Golgi complex~\cite{lee-cell-1988}.
Moreover, spherical GUVs are a canonical tool of in vitro studies.
GUVs are often used to probe static membrane properties, such as the bending modulus and base state surface tension \cite{pecreaux-bassereau-epje-2004, dahl-sm-2016}, as well as dynamic properties, such as the membrane's response to a shear flow \cite{mellema-pre-1997, kantsler-prl-2005, viallat-sm-2008, deschamps-prl-2009}.

\begin{figure}[!b]
	\centering
	\subfigure[\ unperturbed: $\bmx_\sz (\theta, \varphi)$]{
		\hspace{10pt}
		\includegraphics[width=0.36\linewidth]{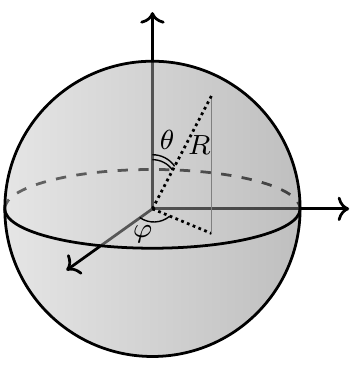}
		\hspace{5pt}
		\label{fig:fig_sphere_unperturbed}
	}
	\subfigure[\ perturbed: $\bmx (\theta, \varphi, t)$]{
		\hspace{10pt}
		\includegraphics[width=0.36\linewidth]{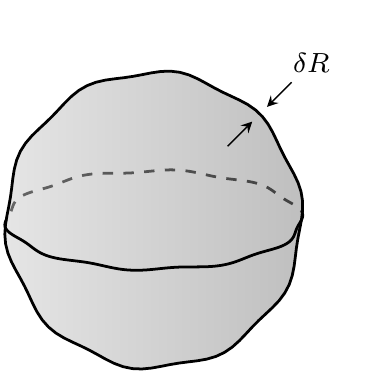}
		\hspace{5pt}
		\label{fig:fig_sphere_perturbed}
	}
	\caption{
		Schematic of the unperturbed (a) and perturbed (b) spherical geometries.
		The sphere has radius $R$ and is characterized by the polar angle $\theta$ and azimuthal angle $\varphi$.
		Membrane perturbations are of characteristic size $\delta R$, with
		$\epsilon := \delta R / R \ll 1$.
	}
	\vspace{-6pt}
	\label{fig:fig_sphere}
\end{figure}

The position of an initially unperturbed membrane vesicle of radius $R$ is given by
\begin{equation} \label{eq:sphere_unperturbed_position}
	\bm{x}_\sz (\theta, \varphi)
	\, = \, R \, \bm{e}_r (\theta, \varphi)
	~,
\end{equation}
where $\theta$ is the polar angle and $\varphi$ is the azimuthal angle of a standard spherical coordinate system (see Fig.~\ref{fig:fig_sphere_unperturbed}).
Similar to the flat case, we denote the unperturbed velocity components and surface tension as $v^\theta_\sz$, $v^\varphi_\sz$, $v_\sz$, and $\lambda_\sz$.
Note that $v^\theta_\sz$ and $v^\varphi_\sz$ have units of inverse time, while $v_\sz$ has units of length per time, as per our differential geometric formulation (see SM \cite[Sec.\ III.1]{supplemental} for details).

For a sphere of fixed shape,
$v_\sz = 0$.
In this case, for a general base flow, neither bending terms nor viscous terms arise in the unperturbed shape equation (see Eq.\ (72) of the SM \cite{supplemental})---a surprising result, since we generally expect both when the membrane is curved (see Eq.~\eqref{eq:general_shape_equation}).
We also find inertial terms are negligible in all cases, as show in the SM \cite[Secs.\ III.3, III.4]{supplemental}.
Consequently, the surface tension is a constant given by
$\lambda_\sz = \lambda^{}_0 := p R / 2$,
which sets the surface tension scale in our scaling analysis as
\begin{equation} \label{eq:sphere_unperturbed_surface_tension_scale}
	\Lambda
	:= \lambda^{}_0
	= \dfrac{p R}{2}
	~.
\end{equation}
To avoid excessive algebra, we consider only base states which are either static or rotating with constant angular velocity about a fixed axis, for which
$v^\theta_\sz = 0$
and
$v^\varphi_\sz = v^\varphi_0$,
where $v^\varphi_0$ is either zero or a nonzero constant.
As shown in the SM \cite[Sec.\ III]{supplemental}, these choices are valid solutions of the unperturbed spherical equations.

We next introduce a radial shape perturbation, such that the perturbed membrane position is given by
\begin{equation} \label{eq:sphere_perturbed_position}
	\bmx (\theta, \varphi, t)
	\, = \, \Big[
		R \, + \epsilon \, \tilr (\theta, \varphi, t)
	\Big] \, \bm{e}_r (\theta, \varphi)
	~,
\end{equation}
where the radial perturbation $\epsilon \, \tilr$ is of characteristic size
$\delta R \ll R$,
as shown in Fig.~\ref{fig:fig_sphere_perturbed}.
We define
$\epsilon := \delta R / R$
to be our small parameter, such that $\tilr$ is $O(R)$.
The fundamental membrane unknowns are expanded about the unperturbed base state solution as
\begin{equation} \label{eq:sphere_unknown_expansions}
	\begin{split}
		v^\theta
		&= \epsilon \, \tilv^\theta
		~,
		\qquad \qquad \quad \hspace{2pt}
		v^\varphi
		= v^\varphi_0
		+ \epsilon \, \tilv^\varphi
		~,
		\\[8pt]
		v
		&= \epsilon \, \tilr_{, t}
		~,
		\qquad
		\text{and}
		\qquad
		\lambda
		= \lambda^{}_0
		+ \epsilon \till
		~.
	\end{split}
\end{equation}
As before, quantities with a `tilde' are assumed to be the same order as their unperturbed counterparts, with the small parameter $\epsilon$ capturing the relative magnitude of base and perturbed quantities.

\def\arraystretch{2.0}
\setlength\tabcolsep{10pt}
\begin{table}[b!]
	\begin{center}
		\def~{\hphantom{0}}
		\begin{tabular}{|c|c|c|c|}
			\hline
			\hline
			$
				\thetand
				:= \theta
			$
			&
			$
				\phind
				:=
				\varphi
			$
			&
			$
				\tilrnd
				:=
				\dfrac{\tilr}{R}
				\rule{0mm}{1.9em}
			$
			&
			$
				\tilv^{\theta \nd}
				:= \dfrac{\tilv^\theta}{\Omega}
			$
			\\[7pt]
			\hline
			$
				\tilv^{\varphi \nd}
				:= \dfrac{\tilv^\varphi}{\Omega}
			$
			&
			$
				\till^*
				:= \dfrac{\till}{\Lambda}
				\rule{0mm}{1.9em}
			$
			&
			$
				t^\nd
				:= \dfrac{t}{\tau}
			$
			&
			$
				\Deltand
				:= R^2 \Delta_{\mathrm{s}}
			$
			\\[7pt]
			\hline
			\hline
		\end{tabular}
		\caption{
			Dimensionless definitions for an initially spherical lipid membrane.
			Here, $\Omega$ is a characteristic angular velocity scale and $\tau$ is a characteristic time scale (see Eq.\ \eqref{eq:sphere_initially_static_general_scalings}).
			\vspace{-10pt}
		}
		\label{tab:tab_sphere_dimensionless_definitions}
	\end{center}
\end{table}

%
%

\subsection{Initially Static Spherical Vesicle} \label{sec:sec_sphere_initially_static_membrane}

For an initially static spherical vesicle, the base state angular velocity
$v^\varphi_0 = 0$.
In this case, only the surface tension scale \eqref{eq:sphere_unperturbed_surface_tension_scale} and length scale $R$ are set in the unperturbed state.
We calculate and non-dimensionalize the perturbed equations in the SM \cite[Sec.\ III.3]{supplemental}, with dimensionless quantities specified as in Table~\ref{tab:tab_sphere_dimensionless_definitions} (see also Refs.\ \cite{olla-pa-2000, vlahovska-2016}).
As the sphere is initially isotropic, we assume the scales of the angular velocity perturbations in $\theta$ and $\varphi$ are identical, and are denoted $\Omega$.
The dimensionless perturbed first-order continuity, in-plane $\theta$, in-plane $\varphi$, and shape equations are respectively given by
\begin{align}
	\tilv^{\theta \nd}_{, \thetand}
	\, + \, \tilv^{\varphi \nd}_{, \phind}
	\, + \, \cot \thetand \, \tilv^{\theta \nd}
	\, + \, 2 \, \tilrnd_{, \tnd}
	\, &= \, 0
	~,
	\label{eq:sphere_perturbed_continuity_static}
	\\[9pt]
	\begin{split}
		\tilv^{\theta \nd}
		+ \, \tilv^{\theta \nd}_{, \thetand \! \thetand}
		+ \, \csc^2 \thetand \tilv^{\theta \nd}_{, \phind \! \phind}
		+ \, \cot \thetand \tilv^{\theta \nd}_{, \thetand}&
		\\
		- \, 2 \cot \thetand \tilv^{\varphi \nd}_{, \phind}
		- \, \cot^2 \thetand \tilv^{\theta \nd}
		+ \, \tillnd_{, \thetand}
		\, &= \, 0
		~,
	\end{split}
	\label{eq:sphere_perturbed_theta_static}
	\\[9pt]
	\begin{split}
		\tilv^{\varphi \nd}_{, \thetand \! \thetand}
		+ \csc^2 \thetand \, \tilv^{\varphi \nd}_{, \phind \! \phind}
		+ \, 2 \cot \thetand \, \csc^2 \thetand \, \tilv^{\theta \nd}_{, \phind}&
		\\
		+ \, 3 \cot \thetand \, \tilv^{\varphi \nd}_{, \thetand}
		+ \, \csc^2 \thetand \, \tillnd_{, \phind}
		\, &= \, 0
		~,
	\end{split}
	\label{eq:sphere_perturbed_phi_static}
	\\
	\intertext{and}
	\Gamma \Big( \!
		2 \, \tilrnd
		+ \Deltand \, \tilrnd
		- 2 \, \tillnd \!
	\Big)
	- \dfrac{1}{2} \Big( \!
		\Deltandd \, \tilrnd
		+ 2 \, \Deltand \tilrnd \!
	\Big)
	&= \, 0
	~.
	\label{eq:sphere_perturbed_shape_static}
\end{align}
In obtaining Eqs.\ \eqref{eq:sphere_perturbed_continuity_static}--\eqref{eq:sphere_perturbed_shape_static}, we find
\begin{equation} \label{eq:sphere_initially_static_general_scalings}
	\Omega
	\, = \, \dfrac{1}{\tau}
	\, = \, \dfrac{\Lambda}{\zeta}
	~,
\end{equation}
such that all terms in the continuity equation \eqref{eq:sphere_perturbed_continuity_static} are the same order, and tension gradients balance viscous forces in the in-plane equations (\ref{eq:sphere_perturbed_theta_static}, \ref{eq:sphere_perturbed_phi_static}).
In this case, the base state surface tension sets the scale of angular velocities and also the time scale over which radial perturbations decay.
In Eq.~\eqref{eq:sphere_perturbed_shape_static} the \FvK\ number $\Gamma$ is given by
\begin{equation} \label{eq:foppl_von_karman_sphere}
	\Gamma
	:= \dfrac{\Lambda R^2}{\kb}
	~,
\end{equation}
and for a sphere the surface Laplacian is defined as
$
	\Deltas (\, \cdot \,)
	:= R^{-2} [
		(\, \cdot \,)_{, \theta \theta}
		+ \cot \theta (\, \cdot \,)_{, \theta}
		+ \csc^2 \theta (\, \cdot \,)_{, \varphi \varphi}
	]
$.

While Eqs.~\eqref{eq:sphere_perturbed_continuity_static}--\eqref{eq:sphere_perturbed_shape_static} contain more terms than their flat counterparts \eqref{eq:flat_static_perturbed_nd_continuity}--\eqref{eq:flat_static_perturbed_nd_shape}, their fundamental structure is similar.
The continuity equation~\eqref{eq:sphere_perturbed_continuity_static} connects in-plane flows with out-of-plane shape deformations, while Eqs.~\eqref{eq:sphere_perturbed_theta_static} and \eqref{eq:sphere_perturbed_phi_static} relate angular velocities and their derivatives to surface tension gradients.
Interestingly, no viscous forces appear in the perturbed shape equation of an initially static vesicle \eqref{eq:sphere_perturbed_shape_static}, as was the case for an initially static flat patch \eqref{eq:flat_static_perturbed_nd_shape}---despite the geometries being different.
The first term in parenthesis in Eq.~\eqref{eq:sphere_perturbed_shape_static} arises from the normal surface tension force $2 \lambda H$, while the second term arises from the bending-induced forces $- 2 \kb H (H^2 - K) - \kb \Deltas H$.
We once again see the \FvK\ number capturing the relative importance of bending and tension terms in governing the membrane's dynamical response to a perturbation.
For example, in GUVs
$\Lambda \sim 10^{-4}$ pN/nm
and
$R \sim 10$ $\mu$m~\cite{pecreaux-bassereau-epje-2004},
while in small membrane vesicles surrounding retrovirus particles
$R \sim 50$ nm~\cite{forster-pnas-2005}.
Assuming
$\Lambda \sim 10^{-3}$ pN/nm
in the latter, $\Gamma$ ranges from $10^{-2}$ to $10^2$, such that the dynamical response of large vesicles is tension dominated while that of small vesicles is bending dominated.
However, as discussed previously, the base state surface tension can span a wide range of values at any radius to enforce areal incompressibility, and $\Gamma$ can span an even wider range of values than those presented here.

%
%

\subsection{Spherical Vesicle with a Base Flow} \label{sec:sec_sphere_base_flow}

When a spherical lipid membrane is placed in a bulk shear flow, the velocity gradient in the surrounding fluid imparts a torque on the membrane and can cause it to rotate about a fixed axis with a nonzero angular velocity
$v^\varphi_0 \ne 0$.
In experimental systems, when the inner and outer fluids are the same viscosity, rotating GUVs are observed in shear flows with shear rates $\dot{\gamma}$ up to $10^{-4}$ $\mu$s$^{-1}$~\cite{mellema-pre-1997, kantsler-prl-2005, deschamps-prl-2009, ota-ac-2009}.
Moreover, in large blood vessels in the human body, shear rates can be as high as
$\dot{\gamma} \sim 10^{-3}$ $\mu$s$^{-1}$ \cite{lipowsky-mr-1980}.
For a spherical lipid membrane vesicle in a shear flow, we assume
$v^\varphi_0 = \dot{\gamma}$
in the base state, and in our non-dimensionalization set the scale of $\tilv^\theta$ and $\tilv^\varphi$ as
\begin{equation} \label{eq:sphere_base_flow_angular_velocity_scale}
	\Omega
	\, := \, v^\varphi_0
	\, = \, \dot{\gamma}
	~.
\end{equation}
Accordingly, we now cannot choose an angular velocity scale to balance in-plane viscous and tension forces.
The base state also sets the surface tension scale $\Lambda$, which once again satisfies Eq.~\eqref{eq:sphere_unperturbed_surface_tension_scale}.

Similar to the planar case, the introduction of a base state angular velocity leads to a new scale over which in-plane quantities vary.
For an initially static sphere, all quantities varied over a length scale $R$.
Equivalently, in-plane quantities were assumed to vary over $O(1)$ changes in the angles $\theta$ and $\varphi$.
In the case of an initially rotating vesicle, similar to the case of a planar membrane with a base flow, such an assumption leads to an unphysical result \cite[Sec.\ III.4.b]{supplemental}.
A scaling analysis reveals there is a new angular scale
\begin{equation} \label{eq:sphere_base_flow_angle_scale}
	\Phi
	:= \dfrac{\zeta \Omega}{\Lambda}
\end{equation}
over which $\tilv^\theta$, $\tilv^\varphi$, and $\till$ vary, while $\tilr$ continues to vary over $O(1)$ changes in $\theta$ and $\varphi$.
This result, which is analogous to that of the planar system, motivates defining the new quantities
\begin{equation} \label{eq:sphere_base_flow_angle_defs}
	\thetandl
	:= \dfrac{\theta}{\Phi}
	\qquad
	\text{and}
	\qquad
	\phindl
	:= \dfrac{\varphi}{\Phi}
\end{equation}
as the angles over which $\tilv^\theta$, $\tilv^\varphi$, and $\till$ vary.
The emergence of a new angular scale in the perturbed equations implies care must be taken when predicting the magnitude of perturbed in-plane and out-of-plane forces.
We also note that in order for $\Phi$ to represent an angle, geometric constraints require
$\Phi \le 1$.

The non-dimensionalization of the linearized, perturbed equations of motion is found in the SM \cite[Sec.\ III.4.b]{supplemental}.
A scaling analysis of the perturbed continuity equation reveals the time scale $\tau$ is given by
\begin{equation} \label{eq:sphere_base_flow_time_scale}
	\tau
	= \dfrac{\zeta}{\Lambda}
	~,
\end{equation}
such that the continuity equation can be written as
\begin{equation} \label{eq:sphere_perturbed_continuity_base_flow}
	\tilv^{\theta \nd}_{, \thetandl}
	\, + \, \tilv^{\varphi \nd}_{, \phindl}
	\, + \, \Phi \big(
		\cot \thetand \, \tilv^{\theta \nd}
		\, + \, 2 \, \tilrnd_{, \phind}
	\big)
	\, + \, 2 \, \tilrnd_{, \tnd}
	\, = \, 0
	~.
\end{equation}
Equation \eqref{eq:sphere_perturbed_continuity_base_flow} connects in-plane flows to out-of-plane shape deformations.
Compared to the perturbed continuity equation of an initially static vesicle~\eqref{eq:sphere_perturbed_continuity_static}, Eq.~\eqref{eq:sphere_perturbed_continuity_base_flow} contains angular derivatives with respect to $\thetandl$ and $\phindl$, and the $\tilrnd_{, \phind}$ term arises from the nonzero base state angular velocity.

Next, the linearized, first-order in-plane equations of motion are given by
\begin{align*}
	\tilv^{\theta \nd}_{, \thetandl \thetandl}
	&+ \csc^2 \thetand \tilv^{\theta \nd}_{, \phindl \phindl}
	+ \Phi \cot \thetand \big(
		\tilv^{\theta \nd}_{, \thetandl}
		- \, 2 \, \tilv^{\varphi \nd}_{, \phindl}
	\big)
	\\
	&+ \, \Phi^2 \, \tilv^{\theta \nd} \big(
		1 - \cot^2 \thetand
	\big)
	+ \, \tillnd_{, \thetandl}
	\, = \, 0
	\stepcounter{equation}
	\tag{\theequation}\label{eq:sphere_perturbed_theta_base_flow}
	\\
	\intertext{and}
	\tilv^{\varphi \nd}_{, \thetandl \thetandl}
	&+ \csc^2 \thetand \, \tilv^{\varphi \nd}_{, \phindl \phindl}
	+ \, \Phi \cot \thetand \big(
		2 \csc^2 \thetand \, \tilv^{\theta \nd}_{, \phindl}
		+ \, 3 \, \tilv^{\varphi \nd}_{, \thetandl}
	\big)
	\\
	&+ \, \csc^2 \thetand \, \tillnd_{, \phindl}
	\, = \, 0
	~,
	\stepcounter{equation}
	\tag{\theequation}\label{eq:sphere_perturbed_phi_base_flow}
\end{align*}
which show the balance between in-plane tension gradients and in-plane viscous forces, and are again similar to their initially static counterparts (\ref{eq:sphere_perturbed_theta_static}, \ref{eq:sphere_perturbed_phi_static}).

Finally, the first-order perturbed shape equation is given by
\begin{align*}
	&2 \, \SL \Big(
		\cos \thetand \big[
			\csc \thetand
			- \sin \thetand
		\big] \tilrnd_{, \thetand \! \phind}
		- \, \cos^2 \thetand \cot^2 \thetand \tilrnd_{, \phind}
	\Big)
	\\[2pt]
	&+ \, \Gamma \Big(
		2 \, \tilrnd
		+ \, \Deltand \, \tilrnd
		- 2 \, \tillnd
	\Big)
	- \dfrac{1}{2} \Big(
		\Deltandd \, \tilrnd
		\, + \, 2 \, \Deltand \, \tilrnd
	\Big)
	= \, 0
	~,
	\stepcounter{equation}
	\tag{\theequation}\label{eq:sphere_perturbed_shape_base_flow}
\end{align*}
where the \FvK\ number is given by Eq.~\eqref{eq:foppl_von_karman_sphere} and the Scriven--Love number is found to be
\begin{equation} \label{eq:sphere_scriven_love}
	\SL
	\, = \, \dfrac{\zeta \Omega R^2}{\kb}
	~.
\end{equation}
The first line in Eq.~\eqref{eq:sphere_perturbed_shape_base_flow} consists of the out-of-plane viscous forces arising from the rotational base flow, which were not present in the perturbed shape equation of a spherical vesicle with no base flow \eqref{eq:sphere_perturbed_shape_static}.
The second line in Eq.~\eqref{eq:sphere_perturbed_shape_base_flow} contains the surface tension and bending forces, which are identical to those found in an initially static sphere (cf.\ Eq.~\eqref{eq:sphere_perturbed_shape_static}).
Note that with Eq.\ \eqref{eq:sphere_scriven_love}, the angular scale $\Phi$ can be expressed as
$\Phi = \Gamma^{-1} \SL$,
such that the ratio of the Scriven--Love and \FvK\ numbers dictates the relative size of various terms in the continuity \eqref{eq:sphere_perturbed_continuity_base_flow} and in-plane (\ref{eq:sphere_perturbed_theta_base_flow}, \ref{eq:sphere_perturbed_phi_base_flow}) equations as well.

For spherical membranes with a base flow, we find three experimentally relevant regimes; in each case, we provide the corresponding symbols in Table~\ref{tab:tab_experiments} and Fig.~\ref{fig:fig_plot}.
First, the bending dominated regime is characteristic of small membrane vesicles, as in the case of 100 nm vesicles surrounding retrovirus particles \cite[$\spadesuit$]{forster-pnas-2005}, for which
$\SL \ll 1$
and
$\Gamma \ll 1$.
In this case, the shape equation \eqref{eq:sphere_perturbed_shape_base_flow} simplifies to contain only the bending terms, and is given by
\begin{equation} \label{eq:sphere_perturbed_shape_base_flow_bending_dominated}
	\Deltandd \, \tilrnd
	\, + \, 2 \, \Deltand \, \tilrnd
	\, = \, 0
	~.
\end{equation}
Next, in large GUVs at low shear rates \cite[$\bowtie$]{mellema-pre-1997}, \cite[$\heartsuit$]{mader-epje-2006}, we find
$\Gamma \gg \SL$
and
$\Gamma \gg 1$,
such that the shape equation is tension dominated and simplifies to
\begin{equation} \label{eq:sphere_perturbed_shape_base_flow_tension_dominated}
	2 \, \tilrnd
	+ \, \Deltand \, \tilrnd
	- 2 \, \tillnd
	\, = \, 0
	~.
\end{equation}
On the other hand, for white blood cells \cite[$\clubsuit$]{kolaczkowska-nri-2013} or GUVs \cite[$\diamondsuit$]{ota-ac-2009} in flows with high shear rates,
$\SL \sim \Gamma \gg 1$
and both viscous and tension forces dominate bending forces.
In this case, the shape equation is given by
\begin{align*}
	&2 \, \SL \, \big(
		\cos \thetand \big[
			\csc \thetand
			- \sin \thetand
		\big] \tilrnd_{, \thetand \! \phind}
		- \cos^2 \thetand \cot^2 \thetand \tilrnd_{, \phind}
	\big)
	\\[4pt]
	&\hspace{7pt}
	+ \Gamma \, \big(
		2 \, \tilrnd
		+ \, \Deltand \, \tilrnd
		- 2 \tillnd
	\big)
	= \, 0
	~.
	\stepcounter{equation}
	\tag{\theequation}\label{eq:sphere_perturbed_shape_base_flow_viscous_tension}
\end{align*}
We refer to systems for which Eq.~\eqref{eq:sphere_perturbed_shape_base_flow_viscous_tension} governs the perturbed out-of-plane dynamics as the viscosity and tension dominated regime.
In such systems, the dynamical response of an initially spherical, rotating lipid membrane vesicle can be significantly affected by the viscous forces arising from the intramembrane fluidity.
In particular, out-of-plane viscous forces could lead to non-trivial corrections in many of the theoretical and numerical studies of membrane-bound vesicles immersed in shearing bulk fluids \cite{keller-skalak-1982, kraus-prl-1996, seifert-epjb-1999, beaucourt-pre-2004, lebedev-prl-2007, noguchi-prl-2007, vlahovska-pre-2007, messlinger-pre-2009, zhao-shaqfeh-jfm-2011, zhao-shaqfeh-pf-2011}.
We reiterate that our analysis does not include effects from the bulk fluid besides the base jump in normal stress, and a comprehensive study involving both the bulk fluid and full membrane equations is necessary to understand vesicle behavior in such situations.

%
%

\section{Cylindrical Membrane Tubes} \label{sec:sec_cylinder}

We lastly consider lipid membrane tubes, which play an important role in many cellular processes, such as material transport between the Golgi complex and the endoplasmic reticulum \cite{lee-cell-1988} and intracellular communication~\cite{rustom-science-2004}.
Tubes are also useful for probing lipid membrane properties, as they can be generated in various ways, including with optical tweezers \cite{dai-bpj-1995, cuvelier-bpj-2005} and molecular motors traveling along microtubules \cite{roux-pnas-2002}.
Moreover, tube pulling is often used to measure the membrane tension and bending rigidity \cite{evans-yeung-cpl-1994, dai-cell-1995, shi-cell-2018}.

As in the previous two sections, we consider lipid membrane tubes which either
(i) are static or
(ii) have a base flow
in their unperturbed state.
Compared to the planar and spherical cases, the cylindrical geometry and its corresponding equations present a new complexity in that the base surface tension scale can be set by either bending forces or the jump in the normal stress across the membrane surface.
Moreover, a tube can have an axial length scale which is much longer than the tube radius, such that quantities can vary over different distances in the axial and angular directions.
The governing equations contain significant differences in the aforementioned scenarios, and are presented systematically in the following sections.

\begin{figure}[!b]
	\centering
	\subfigure[\ unperturbed: $\bmx_\sz (\theta, z)$]{
		\hspace{15pt}
		\includegraphics[width=0.30\linewidth]{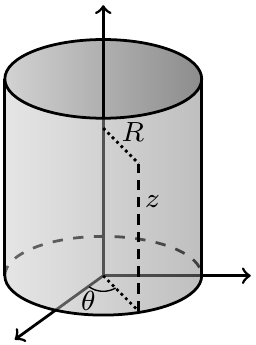}
		\label{fig:fig_cylinder_unperturbed}
		\hspace{5pt}
	}
	\subfigure[\ perturbed: $\bmx (\theta, z, t)$]{
		\hspace{15pt}
		\includegraphics[width=0.30\linewidth]{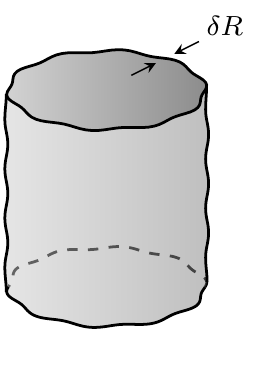}
		\label{fig:fig_cylinder_perturbed}
		\hspace{5pt}
	}
	\caption{
		Schematic of unperturbed (a) and perturbed (b) cylindrical geometries.
		The cylinder has radius $R$ and is characterized by the angle $\theta$ and axial distance $z$.
		Membrane perturbations are of characteristic size $\delta R$, with
		$\epsilon := \delta R / R \ll 1$.
	}
	\label{fig:fig_cylinder}
\end{figure}

The position of an unperturbed cylindrical membrane tube of radius $R$ is given by
\begin{equation} \label{eq:cylinder_unperturbed_position}
	\bm{x}_\sz (\theta, z)
	\, = \, R \, \bm{e}_r (\theta)
	\, + \, z \, \bm{e}_z
	~,
\end{equation}
as shown in Fig.~\ref{fig:fig_cylinder_unperturbed}.
We denote the unperturbed angular velocity, axial velocity, normal velocity, and surface tension as $v^\theta_\sz$, $v^z_\sz$, $v_\sz$, and $\lambda_\sz$, respectively, where $v^\theta_\sz$ has units of radians per time while $v^z_\sz$ and $v_\sz$ have units of length per time.
As the shape is fixed, the normal velocity
$v_\sz = 0$.

The perturbed membrane position is written as
\begin{equation} \label{eq:cylinder_perturbed_position}
	\bm{x} (\theta, z, t)
	\, = \, \Big[
		R
		\, + \, \epsilon \, \tilr (\theta, z, t)
	\Big] \bm{e}_r (\theta)
	\, + \, z \, \bm{e}_z
	~,
\end{equation}
where the small parameter
$\epsilon := \delta R / R \ll 1$
as in the spherical case (see Fig.~\ref{fig:fig_cylinder_perturbed}).
The fundamental membrane unknowns are expanded to first order as
\begin{equation} \label{eq:cylinder_unknown_expansions}
	\begin{split}
		v^\theta
		&= v^\theta_\sz
		+ \epsilon \, \tilv^\theta
		~,
		\qquad \qquad \hspace{5pt}
		v^z
		= v^z_\sz
		+ \epsilon \, \tilv^z
		~,
		\\[4pt]
		v
		&= \epsilon \, \tilr_{, t}
		~,
		\qquad \quad
		\text{and}
		\qquad \quad
		\lambda
		= \lambda_\sz
		+ \epsilon \, \till
		~.
	\end{split}
\end{equation}
As in the planar and spherical cases, quantities with a `tilde' are assumed to be the same order as their unperturbed counterparts.

Before proceeding, we would like to comment on the measurement of membrane tension via tether pulling experiments.
In all cases we consider, the unperturbed shape equation simplifies to
\begin{equation} \label{eq:cylinder_unperturbed_shape}
	\lambda_\sz
	\, = \, \lambda^{}_0
	\, := \, p R
	\, + \, \dfrac{\kb}{4 R^2}
	~,
\end{equation}
where $p$ is the known jump in normal stress across the membrane surface.
Equation \eqref{eq:cylinder_unperturbed_shape} is an extension of the cylindrical Young--Laplace equation,
$\lambda^{}_0 = p R$,
with the addition of nonlinear bending forces which favor a flat membrane.
For a membrane tube with no jump in the normal stress
($p = 0$),
the unperturbed shape equation \eqref{eq:cylinder_unperturbed_shape} implies
$\lambda^{}_0 = \kb / (4 R^2)$
and a force balance shows the pulling force required to hold a static tube is equal to
$f_{\text{pull}} = 2 \pi R (\kb / (4 R^2) + \lambda^{}_0)$ \cite{evans-yeung-cpl-1994, powers-pre-2002, prost-prl-2002}.
The two relations can be combined to show
$f_{\text{pull}} = 2 \pi \sqrt{ \kb \lambda^{}_0 } \,$,
such that by measuring the pulling force via optical tweezers and the bending modulus via independent experiments, one can determine the tension.
This technique is ubiquitous \cite{dai-bpj-1995, cuvelier-bpj-2005, evans-yeung-cpl-1994, dai-cell-1995}, yet as reported in Ref.\ \cite{monnier-prl-2010} it importantly does not hold when there is a jump in the normal stress across the membrane surface.
Such a jump could be caused by both hydrodynamic or osmotic pressure differences, which are generally not measured in the aforementioned experimental studies.
As a result, without prior knowledge of the pressure drop, tension values often cannot be calculated from reported data.
Given these observations, Ref.\ \cite{shi-cell-2018} is the only study we found with sufficient data to approximate the tension scale in membrane tubes; our interpretation of the data is provided in the SM \cite[Secs.\ IV.3.c, IV.4.c]{supplemental}.

\def\arraystretch{2.0}
\setlength\tabcolsep{10pt}
\begin{table}[b!]
	\vspace{10pt}
	\begin{center}
		\def~{\hphantom{0}}
		\begin{tabular}{|c|c|c|c|}
			\hline
			\hline
			$
				\thetand
				:= \theta
			$
			&
			$
				z^\nd
				:= \dfrac{z}{L}
			$
			&
			$
				\tilrnd
				:=
				\dfrac{\tilr}{R}
				\rule{0mm}{1.9em}
			$
			&
			$
				\tilv^{\theta \nd}
				:= \dfrac{\tilv^\theta}{\Omega}
			$
			\\[7pt]
			\hline
			$
				\tilv^{z \nd}
				:= \dfrac{\tilv^z}{V}
			$
			&
			$
				\till^*
				:= \dfrac{\till}{\Lambda}
				\rule{0mm}{1.9em}
			$
			&
			$
				t^\nd
				:= \dfrac{t}{\tau}
			$
			&
			$
				\Deltand
				:= R^2 \Delta_{\mathrm{s}}
			$
			\\[7pt]
			\hline
			\hline
		\end{tabular}
		\caption{
			Dimensionless definitions for an initially cylindrical lipid membrane.
			Here, $\Omega$ is a characteristic angular velocity scale, $V$ is a characteristic axial velocity scale, $L$ is a characteristic axial length scale, and $\tau$ is a characteristic time scale.
			\vspace{-10pt}
		}
		\label{tab:tab_cylinder_dimensionless_definitions}
	\end{center}
\end{table}

%
%

\subsection{Initially Static Membrane Tube} \label{sec:sec_cylinder_initially_static_membrane}

For an initially static cylindrical membrane,
$v^\theta_\sz = 0$,
$v^z_\sz = 0$,
$v_\sz = 0$,
and
$\lambda_\sz = \lambda^{}_0$.
The base state sets the surface tension scale $\Lambda$ as \eqref{eq:cylinder_unperturbed_shape}
\begin{equation} \label{eq:cylinder_static_tension_scale}
	\Lambda
	:= \lambda^{}_0
	= p R
	+ \dfrac{\kb}{4R^2}
	~,
\end{equation}
such that
$
	\lambdand_\sz
	:= \lambda_\sz / \Lambda
	= 1
$.
When
$p \gg \kb / (4 R^3)$,
$\Lambda \approx p R$,
while if
$p \ll \kb / (4 R^3)$
then
$\Lambda \approx \kb / (4 R^2)$.
We refer to these two limits as the pressure and bending limits, respectively.
Note that we consider only cases where
$p \ge 0$,
and defer the analysis of lipid membrane tubes under compression to a future study.

\begin{figure}[!b]
	\vspace{-4pt}
	\centering
	\subfigure[\ $\delta \sim 1$]{
		\includegraphics[width=0.40\linewidth]{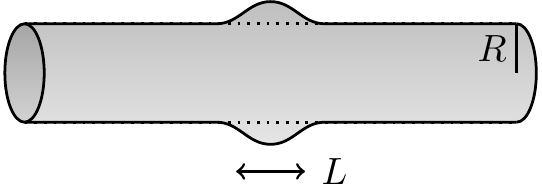}
		\label{fig:fig_delta_unity}
	}
	\hspace{10pt}
	\subfigure[\ $\delta \ll 1$]{
		\includegraphics[width=0.40\linewidth]{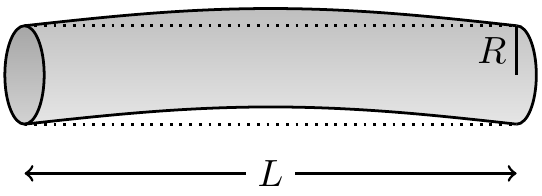}
		\label{fig:fig_delta_small}
	}
	\caption{%
		Schematic showing cylinders of the same aspect ratio, with different values of $\delta$.\
		(a) When
		$\delta \sim 1$,
		out-of-plane quantities vary over a length scale
		$L \sim R$.\
		(b) When
		$\delta \ll 1$,
		out-of-plane quantities vary over a length scale
		$L \gg R$.
		The choice of $\delta$ depends on the membrane behavior under consideration.%
	}
	\label{fig:fig_delta}
\end{figure}

In a tubular lipid membrane, the distance over which quantities vary in the axial direction, $L$, may be much longer than the cylinder radius $R$.
We non-dimensionalize quantities according to the definitions in Table~\ref{tab:tab_cylinder_dimensionless_definitions}, where the angular velocity scale $\Omega$, axial velocity scale $V$, and time scale $\tau$ are to be determined via a scaling analysis.
We also define the parameter $\delta$ to be the ratio of the tube radius to axial length scale, written as
\vspace{-6pt}
\begin{equation} \label{eq:cylinder_delta}
	\delta
	\, := \, \dfrac{R}{L}
	~.
\end{equation}
Note that $\delta$ is not the aspect ratio of the tube, as $L$ is not the cylinder length but rather the axial length scale over which perturbed quantities are expected to vary.
As shown in Fig.\ \ref{fig:fig_delta}, two tubes with the same aspect ratio can have different values of $\delta$, depending on the membrane perturbation and the phenomena of interest.
We now consider separately the case where
$\delta \sim 1$,
which is referred to as a thick tube,
and the case where
$\delta \ll 1$,
from now on referred to as a thin tube.
In both cases, viscous forces enter the perturbed equations and the Scriven--Love number emerges---unlike the initially static flat and spherical geometries.
This result shows the response of an initially static membrane to perturbations is geometry dependent, thus revealing the important relationship between the geometry and dynamics of lipid membranes.

%
%

\subsubsection{Thick Membrane Tube} \label{sec:sec_cylinder_initially_static_thick}

When
$\delta \sim 1$,
the length scales over which perturbed quantities vary in the axial and angular directions are comparable.
In this case, given the definitions in Table~\ref{tab:tab_cylinder_dimensionless_definitions}, the dimensionless first-order perturbed governing equations for a thick, initially static membrane tube are given by (see SM \cite[Sec.\ IV.3.a]{supplemental} and Refs.\ \cite{rahimi-soft-matter-2013, narsimhan-shaqfeh-jfm-2015})
\begin{gather}
	\tilv^{\theta \nd}_{, \thetand}
	\, + \, \tilv^{z \nd}_{, \znd}
	\, + \, \tilrnd_{, \tnd}
	\, = \, 0
	~,
	\label{eq:cylinder_perturbed_continuity_initially_static_thick_nd}
	\\[9pt]
	\tilrnd_{, \tnd \! \thetand}
	\, + \, \tilv^{\theta \nd}_{, \thetand \! \thetand}
	\, + \, \tilv^{\theta \nd}_{, \znd \! \znd}
	\, + \, \tillnd_{, \thetand}
	\, = \, 0
	~,
	\label{eq:cylinder_perturbed_theta_in_plane_initially_static_thick_nd}
	\\[9pt]
	- \tilrnd_{, \tnd \! \znd}
	\, + \, \tilv^{z \nd}_{, \thetand \! \thetand}
	\, + \, \tilv^{z \nd}_{, \znd \! \znd}
	\, + \, \tillnd_{, \znd}
	\, = \, 0
	~,
	\label{eq:cylinder_perturbed_z_in_plane_initially_static_thick_nd}
\end{gather}
and
\begin{equation} \label{eq:cylinder_perturbed_shape_initially_static_thick_nd}
	\begin{split}
		&2 \, \SL \, \tilv^{z \nd}_{, \znd}
		\, + \, \Gamma \big(
			\tilrnd
			+ \, \Deltand \, \tilrnd
			- \, \tillnd
		\big)
		\\[3pt]
		\, &\hspace{7pt}
		- \, \dfrac{1}{4} \, \big(
			3 \, \tilrnd
			+ \, 4 \, \tilrnd_{, \thetand \! \thetand}
			+ \, \Deltand \, \tilrnd
			+ \, 2 \, \Deltandd \, \tilrnd
		\big)
		\, = \, 0
		~.
	\end{split}
\end{equation}
In Eq.~\eqref{eq:cylinder_perturbed_shape_initially_static_thick_nd}, the Scriven--Love and \FvK\ numbers are given by
\begin{equation} \label{eq:cylinder_scriven_love}
	\SL
	\, = \, \dfrac{\zeta V R}{\kb}
	\qquad
	\text{and}
	\qquad
	\Gamma
	\, = \, \dfrac{\Lambda R^2}{\kb}
	~,
\end{equation}
and for scalar quantities the surface Laplacian is given by
$
	\Deltas \, (\, \cdot \,)
	= R^{-2} \, (\, \cdot \,)_{, \theta \theta}
	+ (\, \cdot \,)_{, z z}
$.

A scaling analysis of the perturbed governing equations reveals the time and velocity scales are given by
\begin{equation} \label{eq:cylinder_initially_static_general_scalings}
	\tau
	\, = \, \dfrac{\zeta}{\Lambda}
	~,
	\qquad
	\Omega
	\, = \, \dfrac{\Lambda}{\zeta}
	~,
	\qquad
	\text{and}
	\quad \qquad
	V
	\, = \, \dfrac{R \Lambda}{\zeta}
	~.
\end{equation}
Accordingly, the base state surface tension sets the time scale as well as the scale of in-plane axial and angular velocities.
For lipid membrane tubes with a given geometry, those with larger jumps in the normal stress $p$ in the base state have a larger base state tension, faster perturbed in-plane flows, and more rapid out-of-plane shape rearrangements as well.
Substituting the form of $V$~\eqref{eq:cylinder_initially_static_general_scalings}$_3$ into the Scriven--Love number~\eqref{eq:cylinder_scriven_love}$_1$, we find
$\SL = \Gamma$.

When
$p \ll \kb / (4 R^3)$,
the unperturbed shape equation indicates
$\Lambda \approx \kb / (4 R^2)$.
Consequently,
$\SL = \Gamma = 1/4$
such that surface tension, bending, and viscous forces are all balanced in the perturbed shape equation \eqref{eq:cylinder_perturbed_shape_initially_static_thick_nd}, which simplifies to
\begin{equation} \label{eq:cylinder_perturbed_shape_initially_static_bending_thick_nd}
	\begin{split}
		2 \, \tilv^{z \nd}_{, \znd}
		&+ \, \tilrnd
		+ \, \Deltand \, \tilrnd
		- \, \tillnd
		\\[2pt]
		\, &\hspace{-5pt}
		- \big(
			3 \, \tilrnd
			+ \, 4 \, \tilrnd_{, \thetand \! \thetand}
			+ \, \Deltand \, \tilrnd
			+ \, 2 \, \Deltandd \, \tilrnd
		\big)
		= \, 0
		~.
	\end{split}
\end{equation}
Experimentally, we find several scenarios in this regime \cite[$\P$, $\circledast$]{shi-cell-2018}.

When
$p \gg \kb / (4 R^3)$,
on the other hand, Eq.\ \eqref{eq:cylinder_static_tension_scale} shows the surface tension scales as
$\Lambda \approx p R \gg \kb / (4 R^2)$,
such that
$\SL = \Gamma \gg 1$.
The viscous and tension terms then dominate the bending forces in Eq.~\eqref{eq:cylinder_perturbed_shape_initially_static_thick_nd}, for which the second line is negligible and the shape equation simplifies to
\begin{equation} \label{eq:cylinder_perturbed_shape_initially_static_pressure_thick_nd}
	2 \, \tilv^{z \nd}_{, \znd}
	\, + \, \tilrnd
	\, + \, \Deltand \, \tilrnd
	\, - \, \tillnd
	\, = \, 0
	~.
\end{equation}
Experimentally, we found one study in this regime, where
$\SL = \Gamma \sim 7$ \cite[$\maltese$]{shi-cell-2018}
and bending forces are small relative to tension and viscous forces (see Sec.\ IV.3.c of the SM \cite{supplemental} for experimental details).
In such situations, the equations governing the dynamics of a membrane tube (\ref{eq:cylinder_perturbed_continuity_initially_static_thick_nd}--\ref{eq:cylinder_perturbed_z_in_plane_initially_static_thick_nd}, \ref{eq:cylinder_perturbed_shape_initially_static_pressure_thick_nd}) are identical to those describing a cylindrical, two-dimensional viscous fluid film---for example, a soap bubble \cite[Appendix A.3]{sahu-mandadapu-ale-i}.
Such films, with no bending modulus, are known to undergo a pearling-like instability mediated by in-plane flows when their length exceeds their circumference \cite{sahu-mandadapu-ale-i}.
Therefore, Eqs.\ (\ref{eq:cylinder_perturbed_continuity_initially_static_thick_nd}--\ref{eq:cylinder_perturbed_z_in_plane_initially_static_thick_nd}, \ref{eq:cylinder_perturbed_shape_initially_static_pressure_thick_nd}) indicate lipid membrane tubes with a large stress jump across their surface, for which  $\Gamma \gg 1$ (see Eq.\ \eqref{eq:cylinder_static_tension_scale}), are unstable.
This conclusion is supported by previous studies, which found lipid membrane tubes undergo a pearling instability when
$\Gamma \ge 3/4$ \cite{bar-ziv-pnas-1999, boedec-jfm-2014, narsimhan-shaqfeh-jfm-2015}.

%
%

\subsubsection{Thin Membrane Tube} \label{sec:sec_cylinder_initially_static_thin}

When the length scale $L$ over which axial gradients are expected to occur is much larger than the tube radius $R$,
$\delta \ll 1$
(see Fig.\ \ref{fig:fig_delta_small}), as is the case for membrane tubes found in the endoplasmic reticulum \cite{terasaki-jcb-1986, lippencott-science-2016}.
The dimensionless first-order perturbed governing equations in this case are given by
\begin{gather}
	\tilv^{\theta \nd}_{, \thetand}
	\, + \, \tilv^{z \nd}_{, \znd}
	\, + \, \tilrnd_{, \tnd}
	\, = \, 0
	~,
	\label{eq:cylinder_perturbed_continuity_initially_static_thin_nd}
	\\[9pt]
	\tillnd_{, \thetand}
	\, = \, 0
	~,
	\label{eq:cylinder_perturbed_theta_in_plane_initially_static_thin_nd}
	\\[9pt]
	\tilv^{z \nd}_{, \thetand \! \thetand}
	\, = \, 0
	~,
	\label{eq:cylinder_perturbed_z_in_plane_initially_static_thin_nd}
\end{gather}
and
\begin{equation} \label{eq:cylinder_perturbed_shape_initially_static_thin_nd}
	\begin{split}
		&2 \, \SL \, \delta \, \tilv^{z \nd}_{, \znd}
		\, + \, \Gamma \big(
			\tilrnd
			+ \, \tilrnd_{, \thetand \! \thetand}
			- \, \tillnd
		\big)
		\\
		\, &\hspace{5pt}
		- \, \dfrac{1}{4} \, \big(
			3 \, \tilrnd
			+ \, 5 \, \tilrnd_{, \thetand \! \thetand}
			+ \, 2 \, \tilrnd_{, \thetand \! \thetand \! \thetand \! \thetand}
		\big)
		\, = \, 0
		~.
	\end{split}
\end{equation}
Equations \eqref{eq:cylinder_perturbed_theta_in_plane_initially_static_thin_nd} and \eqref{eq:cylinder_perturbed_z_in_plane_initially_static_thin_nd}, combined with the periodicity of the system, imply $\tillnd$ and $\tilv^{z \nd}$ are both axisymmetric, and can be written as
$\tillnd = \tillnd (\znd, \tnd)$
and
$\tilv^{z \nd} = \tilv^{z \nd} (\znd, \tnd)$.
The time scale $\tau$, angular velocity scale $\Omega$, and axial velocity scale $V$ are found to be (see SM \cite[Sec.\ IV.3.b]{supplemental} for details)
\begin{equation} \label{eq:cylinder_initially_static_general_scalings_thin}
	\tau
	\, = \, \dfrac{\zeta}{\Lambda}
	~,
	\qquad
	\Omega
	\, = \, \dfrac{\Lambda}{\zeta}
	~,
	\qquad
	\text{and}
	\qquad
	V
	\, = \, \dfrac{L \Lambda}{\zeta}
	~,
\end{equation}
and the Scriven--Love and \FvK\ numbers are once again given by Eq.~\eqref{eq:cylinder_scriven_love}.
Importantly, the axial velocity $V$ scales with $L$ such that all terms balance in the continuity equation \eqref{eq:cylinder_perturbed_continuity_initially_static_thin_nd}.
As a result, the factor of $\SL \, \delta$ in the shape equation \eqref{eq:cylinder_perturbed_shape_initially_static_thin_nd} does not vanish when
$\delta \ll 1$,
but rather satisfies
$\SL \, \delta = \Gamma$,
such that viscous and tension forces in the perturbed shape equation are of the same order.

When
$p \ll \kb / (4 R^3)$
and
$\Lambda \approx \kb / (4 R^2)$
in the bending limit (cf.\ Eq.\ \eqref{eq:cylinder_static_tension_scale}),
$\SL \, \delta = \Gamma = 1/4$
and the shape equation \eqref{eq:cylinder_perturbed_shape_initially_static_thin_nd} simplifies to
\begin{equation} \label{eq:cylinder_perturbed_shape_initially_static_thin_nd_bending}
	\begin{split}
		2 \, \tilv^{z \nd}_{, \znd}
		&+ \, \tilrnd
		+ \, \tilrnd_{, \thetand \! \thetand}
		- \, \tillnd
		\\
		\, &- \, \big(
			3 \, \tilrnd
			+ \, 5 \, \tilrnd_{, \thetand \! \thetand}
			+ \, 2 \, \tilrnd_{, \thetand \! \thetand \! \thetand \! \thetand}
		\big)
		\, = \, 0
		~.
	\end{split}
\end{equation}
In the pressure limit where
$p \gg \kb / (4 R^3)$
and
$\Lambda \approx p R$ \eqref{eq:cylinder_static_tension_scale},
$\SL \, \delta = \Gamma \gg 1$
and the shape equation \eqref{eq:cylinder_perturbed_shape_initially_static_thin_nd} simplifies to
\begin{equation} \label{eq:cylinder_perturbed_shape_initially_static_thin_nd_pressure}
	2 \, \tilv^{z \nd}_{, \znd}
	+ \, \tilrnd
	+ \, \tilrnd_{, \thetand \! \thetand}
	- \, \tillnd
	\, = \, 0
	~.
\end{equation}
Thus, for both thick and thin initially static tubes, viscous and tension forces always play an important role in the membrane's dynamical response to perturbations.

In comparing the thick and thin tube equations, for the same base state surface tension, we note several important differences.
First, the thin tube axial velocity scale is larger than its thick tube counterpart by a factor of $\delta^{-1}$, despite the angular velocity and time scales being identical.
Second, thin tubes have axisymmetric axial velocities and surface tensions, while thick tubes in general do not.
Despite these differences, however, thick and thin tubes have identical continuity equations and similar shape equations.
Thus, in both cases radial shape changes lead to axial and in-plane flows, and viscous forces enter the shape equation, leading to the emergence of the Scriven--Love number.

%
%

\subsection{Membrane Tube with a Base Flow} \label{sec:sec_cylinder_base_flow}

Lipid membrane tubes often have a base axial flow, for example when tubes shoot suddenly from the endoplasmic reticulum into the cytoplasm of the cell~\cite{terasaki-jcb-1986}, in neuronal flows along the axon body \cite{dai-bpj-1995, dai-cell-1995}, and in tube pulling experiments with GUVs \cite{roux-pnas-2002} or live cells \cite{upadhyaya-bpj-2004}.
In biological systems, velocities of up to 10 $\mu$m/sec ($10^{-2}$ nm/$\mu$sec) are observed \cite{dai-cell-1995, kaether-mbc-2000}, while velocities of $\le 1$ $\mu$m/sec ($10^{-3}$ nm/$\mu$sec) are more common \cite{leduc-pnas-2004, shi-cell-2018}.
We assume the base state velocity is given by
$\bmv_\sz = v^{}_0 \, \bm{e}_z$,
which sets the velocity scale $V$ as
\begin{equation} \label{eq:cylinder_base_flow_velocity_scale}
	V = v^{}_0
	~.
\end{equation}
In this case, the unperturbed shape equation is given by Eq.\ \eqref{eq:cylinder_unperturbed_shape} and the surface tension scale is again defined as in Eq.\ \eqref{eq:cylinder_static_tension_scale}.

As was the case for planar and spherical lipid membranes, a nontrivial scaling analysis is required to non-dimensionalize the equations governing tubular systems.
As shown in the SM \cite[Sec.\ IV.4.b]{supplemental}, we find
(i) out-of-plane shape perturbations vary over a length $L$ in the axial direction,
(ii) all quantities vary over $O(1)$ changes in the angle $\theta$, and
(iii) the in-plane quantities $\tilv^z$, $\tilv^\theta$, and $\till$ vary over a length scale
\begin{equation} \label{eq:cylinder_base_flow_new_length_scale}
	\ell
	= \dfrac{\zeta V}{\Lambda}
\end{equation}
in the axial direction.
Thus,
$O(\tilv^z_{, z}) = V / \ell$,
$O(\tilv^z_{, \theta}) = V$,
and
$O(\tilr_{, z}) = R / L$.
Due to there being different characteristic lengths in the axial direction, we define the new dimensionless variable
\begin{equation} \label{eq:cylinder_base_flow_new_axial_scaling}
	\zndl
	:= \dfrac{z}{\ell}
	~.
\end{equation}

\begin{figure}[!t]
	\centering
	\includegraphics[width=0.90\linewidth]{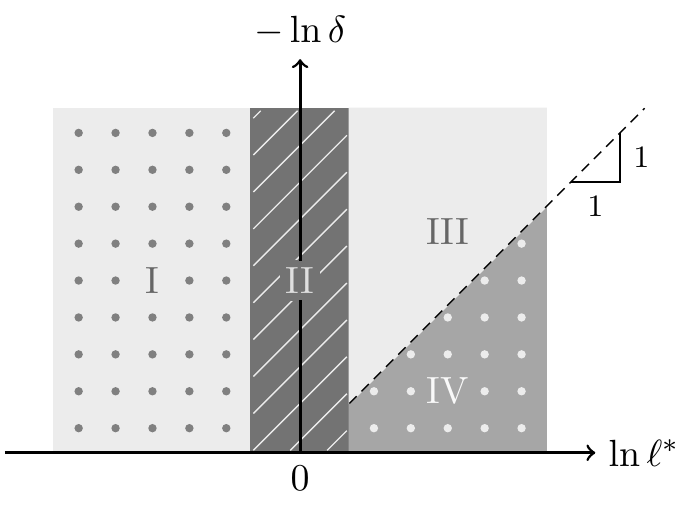}
	\caption{
		Schematic of the cylindrical regimes, which are defined by $\delta$, the ratio of the radius to the axial length scale \eqref{eq:cylinder_delta}, and the parameter $\ellnd$ \eqref{eq:cylinder_base_flow_capillary}, which captures the dimensionless base flow velocity.
		Note the logarithmic scale on both axes.
		Regimes I and II correspond to small ($\ellnd \ll 1$) and moderate ($\ellnd \sim 1$) velocities, respectively.
		For large velocities ($\ellnd \gg 1$), the governing equations differ for thin tubes ($\delta \ll 1$, Regime III) and thick tubes ($\delta \sim 1$, Regime IV).
		The equations governing the dynamics of each regime are provided in the main text.
	}
	\label{fig:fig_cylinder_regimes}
\end{figure}

With the introduction of the length sale $\ell$ \eqref{eq:cylinder_base_flow_new_length_scale}, there are three relevant length scales for the cylinder: $\ell$, $R$, and $L$.
The ratios of these quantities are captured by two dimensionless parameters: the ratio
$\delta = R / L \le 1$ \eqref{eq:cylinder_delta}
and the parameter
\begin{equation} \label{eq:cylinder_base_flow_capillary}
	\ellnd
	\, := \, \dfrac{\ell}{R}
	\,  = \, \dfrac{\zeta V}{\Lambda R}
	~.
\end{equation}
As depicted in Fig.~\ref{fig:fig_cylinder_regimes}, the values of the parameters $\delta$ and $\ellnd$ lead to four regimes with different governing equations, which are considered separately in the subsequent sections.
However, before discussing each regime individually, we first highlight their commonalities, with the non-dimensionalization of the governing equations presented in Sec.\ IV.4 of the SM \cite{supplemental}.

First, in all cases the Scriven--Love and \FvK\ numbers are again given by Eq.\ \eqref{eq:cylinder_scriven_love}, and set the dimensionless parameter $\ellnd$ according to
$\ellnd = \Gamma^{-1} \SL$.
Furthermore, in all regimes the linearized, perturbed shape equation is given by
\begin{align*}
	&\dfrac{2 \SL}{\ellnd} \, \tilv^{z \nd}_{, \zndl}
	+ \, \Gamma \big(
		\tilrnd
		+ \Deltand \, \tilrnd
		- \tillnd
	\big)
	\stepcounter{equation}
	\tag{\theequation}\label{eq:cylinder_base_flow_perturbed_shape}
	\\
	&\hspace{5pt}
	- \dfrac{1}{4} \big(
		3 \tilrnd
		+ \, 4 \tilrnd_{, \thetand \! \thetand}
		+ \, \Deltand \, \tilrnd
		+ \, 2 \, \Deltandd \, \tilrnd
	\big)
	= \, 0
	~,
\end{align*}
where although
$\SL / \ellnd = \Gamma$,
we include both numbers to delineate viscous and tension forces.
As before, we consider only cases where the jump in the normal stress
$p \ge 0$,
for which
$\Gamma \ge 1/4$
and
$\SL / \ellnd \ge 1/4$---indicating viscous forces in the normal direction are always significant, irrespective of the speed of the base flow.
Additionally, in the case where
$p \gg \kb / (4 R^3)$ \eqref{eq:cylinder_static_tension_scale},
the surface tension scales as
$\Lambda \approx p R \gg \kb / (4 R^2)$,
such that
$\SL / \ellnd = \Gamma \gg 1$
and bending forces are negligible compared to viscous forces and tension forces in the normal direction.
In the limit of a large normal stress jump in the base state, the shape equation \eqref{eq:cylinder_base_flow_perturbed_shape} again reduces to that of a two-dimensional fluid film with a base flow.
We previously found such films could admit time-oscillating solutions, unlike their initially static counterparts, and also undergo a pearling instability \cite{sahu-mandadapu-ale-i}.
The characterization of the instabilities of lipid membrane tubes, with and without a base flow, is the subject of a future study~\cite{tchoufag-convective}.

In each of the four regimes shown in Fig.\ \ref{fig:fig_cylinder_regimes} and characterized by the values of $\delta$ and $\ellnd$, the shape equation is given by Eq.~\eqref{eq:cylinder_base_flow_perturbed_shape}.
In what follows, we provide the continuity and in-plane equations in each case.
The non-dimensionalization of all equations is provided in the SM \cite[Sec.\ IV.4]{supplemental}.
We note that for a tube of a given radius $R$ and normal stress jump $p$ in the base state, the surface tension scale is set according to Eq.\ \eqref{eq:cylinder_static_tension_scale}.
Consequently, $\ellnd$ captures the ratio of the base state velocity $V$ to the velocity scale in the absence of a base flow, $R \Lambda / \zeta$ \eqref{eq:cylinder_initially_static_general_scalings}$_3$.

%
%

\subsubsection*{\texorpdfstring{Regime I: $\ellnd \ll 1$}{Regime I}} \label{sec:sec_cylinder_base_flow_regime_1}

In the first regime, the base velocity $V$ is small relative to the intrinsic velocity scale $R \Lambda / \zeta$ such that
$\ellnd \ll 1$,
as is the case in the tube pulling experiments of Ref.\ \cite[$\twonotes$]{shi-cell-2018} where
$\ellnd \sim 2 \cdot 10^{-3}$.
In this case, the dimensionless first-order perturbed continuity, in-plane $\theta$, and in-plane $z$ equations simplify to
\begin{gather}
	\tilv^{\theta \nd}_{, \thetand}
	+ \, \tilv^{z \nd}_{, \zndl}
	+ \, \tilrnd_{, \tnd}
	= \, 0
	~,
	\label{eq:cylinder_base_flow_perturbed_continuity_regime_1}
	\\[10pt]
	\tilv^{\theta \nd}_{, \zndl \zndl}
	= \, 0
	~,
	\label{eq:cylinder_base_flow_perturbed_in_plane_theta_regime_1}
	\\
	\intertext{and}
	\tilv^{z \nd}_{, \zndl \zndl}
	+ \, \tillnd_{, \zndl}
	= \, 0
	~.
	\label{eq:cylinder_base_flow_perturbed_in_plane_z_regime_1}
\end{gather}
The angular velocities are at most linear in $\zndl$ \eqref{eq:cylinder_base_flow_perturbed_in_plane_theta_regime_1} and axial surface tension changes are balanced by two axial derivatives of the $z$-velocity \eqref{eq:cylinder_base_flow_perturbed_in_plane_z_regime_1}.
The angular velocity scale and time scale are respectively given by
\begin{equation} \label{eq:cylinder_base_flow_scales}
	\Omega
	\, = \, \dfrac{\Lambda}{\zeta}
	\qquad
	\text{and}
	\qquad
	\tau
	\, = \, \dfrac{\zeta}{\Lambda}
	~,
\end{equation}
as is the case for Regimes II and III as well.

%
%

\subsubsection*{\texorpdfstring{Regime II: $\ellnd \sim 1$}{Regime II}} \label{sec:sec_cylinder_base_flow_regime_2}

When the base velocity scale $V$ is comparable to the intrinsic velocity scale $R \Lambda / \zeta$, the length scale $\ell$ is comparable to the radius $R$.
In this case, the linearized continuity, in-plane $\theta$, and in-plane $z$ equations are given by
\begin{gather}
	\tilv^{\theta \nd}_{, \thetand}
	+ \, \tilv^{z \nd}_{, \zndl}
	\, + \, \delta \, \tilrnd_{, \znd}
	+ \, \tilrnd_{, \tnd}
	= \, 0
	~,
	\label{eq:cylinder_base_flow_perturbed_continuity_regime_2}
	\\[10pt]
	\tilrnd_{, \tnd \! \thetand}
	+ \, \tilv^{\theta \nd}_{, \thetand \! \thetand}
	+ \, \tilv^{\theta \nd}_{, \zndl \zndl}
	\, + \, \delta \, \tilrnd_{, \thetand \! \znd}
	+ \, \tillnd_{, \thetand}
	= \, 0
	~,
	\label{eq:cylinder_base_flow_perturbed_in_plane_theta_regime_2}
	\\
	\intertext{and}
	- \delta \, \tilrnd_{, \tnd \! \znd}
	+ \tilv^{z \nd}_{, \thetand \! \thetand}
	+ \, \tilv^{z \nd}_{, \zndl \zndl}
	\, - \, \delta^2 \, \tilrnd_{, \znd \! \znd}
	+ \, \till_{, \zndl}
	= \, 0
	~,
	\label{eq:cylinder_base_flow_perturbed_in_plane_z_regime_2}
\end{gather}
where for long, thin tubes
($\delta \ll 1$)
all terms containing factors of $\delta$ are negligible.

%
%

\subsubsection*{\texorpdfstring{Regime III: $1 \ll \ellnd \ll \delta^{-1}$}{Regime III}} \label{sec:sec_cylinder_base_flow_regime_3}

For thin tubes with high base flow velocities, the linearized perturbed governing equations are given by
\begin{gather}
	\tilv^{\theta \nd}_{, \thetand}
	+ \, \tilv^{z \nd}_{, \zndl}
	+ \, \tilrnd_{, \tnd}
	= \, 0
	~,
	\label{eq:cylinder_base_flow_perturbed_continuity_regime_3}
	\\[10pt]
	\tillnd_{, \thetand}
	= \, 0
	~,
	\label{eq:cylinder_base_flow_perturbed_in_plane_theta_regime_3}
	\\
	\intertext{and}
	\tilv^{z \nd}_{, \thetand \! \thetand}
	= \, 0
	~.
	\label{eq:cylinder_base_flow_perturbed_in_plane_z_regime_3}
\end{gather}
Similar to the initially static thin tube discussed in Sec.\ \ref{sec:sec_cylinder_initially_static_thin}, in this regime both $\tilv^{z \nd}$ and $\tillnd$ are axisymmetric (see SM \cite[Sec.\ IV.4.d]{supplemental}).
Comparing Eqs.\ \eqref{eq:cylinder_base_flow_perturbed_continuity_regime_3}--\eqref{eq:cylinder_base_flow_perturbed_in_plane_z_regime_3} to Eqs.\ \eqref{eq:cylinder_perturbed_continuity_initially_static_thin_nd}--\eqref{eq:cylinder_perturbed_z_in_plane_initially_static_thin_nd} in the aforementioned section, the only difference is the $\tilv^{z \nd}_{, \zndl}$ term in Eq.\ \eqref{eq:cylinder_base_flow_perturbed_continuity_regime_3} replaces $\tilv^{z \nd}_{, \znd}$ in Eq.~\eqref{eq:cylinder_perturbed_continuity_initially_static_thin_nd}.

%
%

\subsubsection*{\texorpdfstring{Regime IV: $1 \le \delta^{-1} \ll \ellnd$}{Regime IV}} \label{sec:sec_cylinder_base_flow_regime_4}

For thicker tubes at high velocities, the length scale $\ell$ \eqref{eq:cylinder_base_flow_new_length_scale} can become longer than the axial length scale $L$.
This is the only regime in which
$\ell > L$,
and the characteristic angular velocity and time scales must be rescaled: they are found to be (see Sec.\ IV.4.c of the SM \cite{supplemental})
\begin{equation} \label{eq:cylinder_base_flow_scales_long}
	\Omega
	\, = \, \dfrac{V}{L}
	\qquad
	\text{and}
	\qquad
	\tau
	\, = \, \dfrac{L}{V}
	~.
\end{equation}
The dimensionless perturbed first-order continuity, in-plane $\theta$, and in-plane $z$ equations are given by
\begin{gather}
	\tilv^{\theta \nd}_{, \thetand}
	+ \, \tilrnd_{, \znd}
	+ \, \tilrnd_{, \tnd}
	= \, 0
	~,
	\label{eq:cylinder_base_flow_perturbed_continuity_regime_4}
	\\[10pt]
	- \tilv^{z \nd}_{, \zndl \thetand}
	+ \, \tillnd_{, \thetand}
	= \, 0
	~,
	\label{eq:cylinder_base_flow_perturbed_in_plane_theta_regime_4}
	\\
	\intertext{and}
	- \delta^2 \, \tilrnd_{, \tnd \! \znd}
	+ \tilv^{z \nd}_{, \thetand \! \thetand}
	- \delta^2 \, \tilrnd_{, \znd \! \znd}
	= \, 0
	~.
	\label{eq:cylinder_base_flow_perturbed_in_plane_z_regime_4}
\end{gather}
When
$\delta \ll 1$,
Eq.\ \eqref{eq:cylinder_base_flow_perturbed_in_plane_z_regime_4} simplifies to
$\tilv^{z \nd}_{, \thetand \! \thetand} = 0$,
again implying
$\tilv^{z \nd} = \tilv^{z \nd} (\zndl, \tnd)$
and
$\tillnd = \tillnd(\zndl, \tnd)$.

With the governing equations in each of the four regimes, we recognize the importance of a scaling analysis in elucidating the relative magnitude of various in-plane and out-of-plane forces.
Our analysis in this section closes our discussion of lipid membrane tubes, with and without a base flow.
We highlight that in all cases considered, the Scriven--Love and \FvK\ numbers are given by Eq.\ \eqref{eq:cylinder_scriven_love}.
Moreover, in every situation, viscous and tension forces are found to be significant in describing the dynamics of a perturbed lipid membrane tube, as also shown in Table \ref{tab:tab_experiments} and Fig.\ \ref{fig:fig_plot}.

%
%

\section{Conclusion} \label{sec:sec_conclusion}

In this paper, we determined and non-dimensionalized the linearized equations of lipid membranes perturbed about three commonly occurring geometries: flat patches, spherical vesicles, and cylindrical tubes.
We (i) found perturbed in-plane quantities vary over a new length scale when there is a base flow, and (ii) also found a new dimensionless number, the Scriven--Love number $\SL$, which compares out-of-plane forces arising from the in-plane intramembrane viscosity to the well-known out-of-plane bending forces.
For each of the three geometries, we analyzed relevant experiments involving lipid and biological membranes.

Though to our knowledge
$\SL \ll 1$
in all experiments involving flat membranes, we found biologically relevant situations in which
$\SL \sim 1$,
and even
$\SL \gg 1$,
in perturbed spheres and cylinders.
In this manner, we demonstrated that the in-plane viscous flow of lipids cannot be ignored when understanding lipid membrane dynamics in general geometric configurations.
Our calculation of the Scriven--Love number in a variety of experimental studies (Table~\ref{tab:tab_experiments} and Fig.~\ref{fig:fig_plot}) shows the in-plane intramembrane fluidity is significant in many biological settings, and emphasizes the importance of measuring characteristic velocities and surface tensions in experimental systems---which are currently often not reported.
Moreover, we found that different terms arose in the equations of motion of different geometries.
For example, perturbed, initially static lipid membrane tubes are acted upon by out-of-plane viscous forces, while the corresponding flat patches and spherical vesicles are not.
Accordingly, geometry plays an important role in understanding the dynamics of lipid membranes.

Throughout this work, we assumed a constant positive jump in the normal stress $p$ acting on the membrane in its base configuration.
In biological systems, however, lipid membranes are surrounded by fluid on both sides and feel body forces due to the bulk fluid stresses acting on the membrane surface.
Thus, as the membrane deforms and displaces the surrounding fluid, the bulk fluid stress will have first-order corrections which then enter the perturbed membrane equations of motion in both the in-plane and out-of-plane directions.
Including higher-order effects from the bulk fluid would be a natural extension of our work.

%
%

\begin{acknowledgments}
	We thank Prof.\ Paul Chaikin for stimulating discussions, Prof.\ Howard Stone for insightful comments on the manuscript, Prof.\ Petia Vlahovska for helpful remarks, and Dr.\ Patricia Bassereau for bringing Ref.\ \cite{monnier-prl-2010} to our attention.
	A.S.\ acknowledges the support of the Computational Science Graduate Fellowship from the U.S.\ Department of Energy, as well as U.C.\ Berkeley.
	J.T.\ acknowledges the support of U.C.\ Berkeley.
	K.K.M.\ is supported by Department of Energy Contract No.\ DE-AC02-05CH11231, FWP no.\ CHPHYS02.
\end{acknowledgments}

%
%

\bibliographystyle{apsrev4-1}
\bibliography{refs}

\begin{thebibliography}{85}%
\makeatletter
\providecommand \@ifxundefined [1]{%
 \@ifx{#1\undefined}
}%
\providecommand \@ifnum [1]{%
 \ifnum #1\expandafter \@firstoftwo
 \else \expandafter \@secondoftwo
 \fi
}%
\providecommand \@ifx [1]{%
 \ifx #1\expandafter \@firstoftwo
 \else \expandafter \@secondoftwo
 \fi
}%
\providecommand \natexlab [1]{#1}%
\providecommand \enquote  [1]{``#1''}%
\providecommand \bibnamefont  [1]{#1}%
\providecommand \bibfnamefont [1]{#1}%
\providecommand \citenamefont [1]{#1}%
\providecommand \href@noop [0]{\@secondoftwo}%
\providecommand \href [0]{\begingroup \@sanitize@url \@href}%
\providecommand \@href[1]{\@@startlink{#1}\@@href}%
\providecommand \@@href[1]{\endgroup#1\@@endlink}%
\providecommand \@sanitize@url [0]{\catcode `\\12\catcode `\$12\catcode
  `\&12\catcode `\#12\catcode `\^12\catcode `\_12\catcode `\%12\relax}%
\providecommand \@@startlink[1]{}%
\providecommand \@@endlink[0]{}%
\providecommand \url  [0]{\begingroup\@sanitize@url \@url }%
\providecommand \@url [1]{\endgroup\@href {#1}{\urlprefix }}%
\providecommand \urlprefix  [0]{URL }%
\providecommand \Eprint [0]{\href }%
\providecommand \doibase [0]{http://dx.doi.org/}%
\providecommand \selectlanguage [0]{\@gobble}%
\providecommand \bibinfo  [0]{\@secondoftwo}%
\providecommand \bibfield  [0]{\@secondoftwo}%
\providecommand \translation [1]{[#1]}%
\providecommand \BibitemOpen [0]{}%
\providecommand \bibitemStop [0]{}%
\providecommand \bibitemNoStop [0]{.\EOS\space}%
\providecommand \EOS [0]{\spacefactor3000\relax}%
\providecommand \BibitemShut  [1]{\csname bibitem#1\endcsname}%
\let\auto@bib@innerbib\@empty
\bibitem [{\citenamefont {Watanabe}\ \emph
  {et~al.}(2013{\natexlab{a}})\citenamefont {Watanabe}, \citenamefont {Rost},
  \citenamefont {Camacho-P\'{e}rez}, \citenamefont {Davis}, \citenamefont
  {S\"{o}hl-Kielczynski}, \citenamefont {Rosenmund},\ and\ \citenamefont
  {Jorgensen}}]{watanabe-nature-2013}%
  \BibitemOpen
  \bibfield  {author} {\bibinfo {author} {\bibfnamefont {S.}~\bibnamefont
  {Watanabe}}, \bibinfo {author} {\bibfnamefont {B.~R.}\ \bibnamefont {Rost}},
  \bibinfo {author} {\bibfnamefont {M.}~\bibnamefont {Camacho-P\'{e}rez}},
  \bibinfo {author} {\bibfnamefont {M.~W.}\ \bibnamefont {Davis}}, \bibinfo
  {author} {\bibfnamefont {B.}~\bibnamefont {S\"{o}hl-Kielczynski}}, \bibinfo
  {author} {\bibfnamefont {C.}~\bibnamefont {Rosenmund}}, \ and\ \bibinfo
  {author} {\bibfnamefont {E.~M.}\ \bibnamefont {Jorgensen}},\ }\href
  {https://doi.org/10.1038/nature12809} {\bibfield  {journal} {\bibinfo
  {journal} {Nature}\ }\textbf {\bibinfo {volume} {504}},\ \bibinfo {pages}
  {242} (\bibinfo {year} {2013}{\natexlab{a}})}\BibitemShut {NoStop}%
\bibitem [{\citenamefont {Terasaki}\ \emph {et~al.}(1986)\citenamefont
  {Terasaki}, \citenamefont {Chen},\ and\ \citenamefont
  {Fujiwara}}]{terasaki-jcb-1986}%
  \BibitemOpen
  \bibfield  {author} {\bibinfo {author} {\bibfnamefont {M.}~\bibnamefont
  {Terasaki}}, \bibinfo {author} {\bibfnamefont {L.~B.}\ \bibnamefont {Chen}},
  \ and\ \bibinfo {author} {\bibfnamefont {K.}~\bibnamefont {Fujiwara}},\
  }\href {https://dx.doi.org/10.1083/jcb.103.4.1557} {\bibfield  {journal}
  {\bibinfo  {journal} {J. Cell Biol.}\ }\textbf {\bibinfo {volume} {103}},\
  \bibinfo {pages} {1557} (\bibinfo {year} {1986})}\BibitemShut {NoStop}%
\bibitem [{\citenamefont {Nixon-Abell}\ \emph {et~al.}(2016)\citenamefont
  {Nixon-Abell}, \citenamefont {Obara}, \citenamefont {Weigel}, \citenamefont
  {Li}, \citenamefont {Legant}, \citenamefont {Xu}, \citenamefont {Pasolli},
  \citenamefont {Harvey}, \citenamefont {Hess}, \citenamefont {Betzig},
  \citenamefont {Blackstone},\ and\ \citenamefont
  {Lippincott-Schwartz}}]{lippencott-science-2016}%
  \BibitemOpen
  \bibfield  {author} {\bibinfo {author} {\bibfnamefont {J.}~\bibnamefont
  {Nixon-Abell}}, \bibinfo {author} {\bibfnamefont {C.~J.}\ \bibnamefont
  {Obara}}, \bibinfo {author} {\bibfnamefont {A.~V.}\ \bibnamefont {Weigel}},
  \bibinfo {author} {\bibfnamefont {D.}~\bibnamefont {Li}}, \bibinfo {author}
  {\bibfnamefont {W.~R.}\ \bibnamefont {Legant}}, \bibinfo {author}
  {\bibfnamefont {C.~S.}\ \bibnamefont {Xu}}, \bibinfo {author} {\bibfnamefont
  {H.~A.}\ \bibnamefont {Pasolli}}, \bibinfo {author} {\bibfnamefont
  {K.}~\bibnamefont {Harvey}}, \bibinfo {author} {\bibfnamefont {H.~F.}\
  \bibnamefont {Hess}}, \bibinfo {author} {\bibfnamefont {E.}~\bibnamefont
  {Betzig}}, \bibinfo {author} {\bibfnamefont {C.}~\bibnamefont {Blackstone}},
  \ and\ \bibinfo {author} {\bibfnamefont {J.}~\bibnamefont
  {Lippincott-Schwartz}},\ }\href {https://dx.doi.org/10.1126/science.aaf3928}
  {\bibfield  {journal} {\bibinfo  {journal} {Science}\ }\textbf {\bibinfo
  {volume} {354}} (\bibinfo {year} {2016})}\BibitemShut {NoStop}%
\bibitem [{\citenamefont {Evans}\ and\ \citenamefont
  {Skalak}(1980)}]{evans-skalak}%
  \BibitemOpen
  \bibfield  {author} {\bibinfo {author} {\bibfnamefont {E.~A.}\ \bibnamefont
  {Evans}}\ and\ \bibinfo {author} {\bibfnamefont {R.}~\bibnamefont {Skalak}},\
  }\href@noop {} {\emph {\bibinfo {title} {Mechanics and Thermodynamics of
  Biomembranes}}}\ (\bibinfo  {publisher} {CRC Press},\ \bibinfo {address}
  {Boca Raton, Fl.},\ \bibinfo {year} {1980})\BibitemShut {NoStop}%
\bibitem [{\citenamefont {Zhong-Can}\ and\ \citenamefont
  {Helfrich}(1989)}]{ou-yang-pra-1989}%
  \BibitemOpen
  \bibfield  {author} {\bibinfo {author} {\bibfnamefont {O.-Y.}\ \bibnamefont
  {Zhong-Can}}\ and\ \bibinfo {author} {\bibfnamefont {W.}~\bibnamefont
  {Helfrich}},\ }\href {https://doi.org/10.1103/PhysRevA.39.5280} {\bibfield
  {journal} {\bibinfo  {journal} {Phys. Rev. A}\ }\textbf {\bibinfo {volume}
  {39}},\ \bibinfo {pages} {5280} (\bibinfo {year} {1989})}\BibitemShut
  {NoStop}%
\bibitem [{\citenamefont {Seifert}\ \emph {et~al.}(1991)\citenamefont
  {Seifert}, \citenamefont {Berndl},\ and\ \citenamefont
  {Lipowsky}}]{seifert-pra-1991}%
  \BibitemOpen
  \bibfield  {author} {\bibinfo {author} {\bibfnamefont {U.}~\bibnamefont
  {Seifert}}, \bibinfo {author} {\bibfnamefont {K.}~\bibnamefont {Berndl}}, \
  and\ \bibinfo {author} {\bibfnamefont {R.}~\bibnamefont {Lipowsky}},\ }\href
  {https://doi.org/10.1103/PhysRevA.44.1182} {\bibfield  {journal} {\bibinfo
  {journal} {Phys. Rev. A}\ }\textbf {\bibinfo {volume} {44}},\ \bibinfo
  {pages} {1182} (\bibinfo {year} {1991})}\BibitemShut {NoStop}%
\bibitem [{\citenamefont {Fournier}(1996)}]{fournier-prl-1996}%
  \BibitemOpen
  \bibfield  {author} {\bibinfo {author} {\bibfnamefont {J.~B.}\ \bibnamefont
  {Fournier}},\ }\href {https://doi.org/10.1103/PhysRevLett.76.4436} {\bibfield
   {journal} {\bibinfo  {journal} {Phys. Rev. Lett.}\ }\textbf {\bibinfo
  {volume} {76}},\ \bibinfo {pages} {4436} (\bibinfo {year}
  {1996})}\BibitemShut {NoStop}%
\bibitem [{\citenamefont {Seifert}(1997)}]{seifert-long}%
  \BibitemOpen
  \bibfield  {author} {\bibinfo {author} {\bibfnamefont {U.}~\bibnamefont
  {Seifert}},\ }\href {https://dx.doi.org/10.1080/00018739700101488} {\bibfield
   {journal} {\bibinfo  {journal} {Adv. Phys.}\ }\textbf {\bibinfo {volume}
  {46}},\ \bibinfo {pages} {13} (\bibinfo {year} {1997})}\BibitemShut {NoStop}%
\bibitem [{\citenamefont {Steigmann}(1999)}]{steigmann-arma-1999}%
  \BibitemOpen
  \bibfield  {author} {\bibinfo {author} {\bibfnamefont {D.~J.}\ \bibnamefont
  {Steigmann}},\ }\href {https://doi.org/10.1007/s002050050183} {\bibfield
  {journal} {\bibinfo  {journal} {Arch. Ration. Mech. Anal.}\ }\textbf
  {\bibinfo {volume} {150}},\ \bibinfo {pages} {127} (\bibinfo {year}
  {1999})}\BibitemShut {NoStop}%
\bibitem [{\citenamefont {Powers}\ \emph {et~al.}(2002)\citenamefont {Powers},
  \citenamefont {Huber},\ and\ \citenamefont {Goldstein}}]{powers-pre-2002}%
  \BibitemOpen
  \bibfield  {author} {\bibinfo {author} {\bibfnamefont {T.~R.}\ \bibnamefont
  {Powers}}, \bibinfo {author} {\bibfnamefont {G.}~\bibnamefont {Huber}}, \
  and\ \bibinfo {author} {\bibfnamefont {R.~E.}\ \bibnamefont {Goldstein}},\
  }\href {https://dx.doi.org/10.1103/PhysRevE.65.041901} {\bibfield  {journal}
  {\bibinfo  {journal} {Phys. Rev. E}\ }\textbf {\bibinfo {volume} {65}},\
  \bibinfo {pages} {041901} (\bibinfo {year} {2002})}\BibitemShut {NoStop}%
\bibitem [{\citenamefont {Capovilla}\ and\ \citenamefont
  {Guven}(2002)}]{capovilla-guven-jpa-2002}%
  \BibitemOpen
  \bibfield  {author} {\bibinfo {author} {\bibfnamefont {R.}~\bibnamefont
  {Capovilla}}\ and\ \bibinfo {author} {\bibfnamefont {J.}~\bibnamefont
  {Guven}},\ }\href {https://doi.org/10.1088/0305-4470/35/30/302} {\bibfield
  {journal} {\bibinfo  {journal} {J. Phys. A: Math. Gen.}\ }\textbf {\bibinfo
  {volume} {35}},\ \bibinfo {pages} {6233} (\bibinfo {year}
  {2002})}\BibitemShut {NoStop}%
\bibitem [{\citenamefont {Guven}(2004)}]{guven-jpa-2004}%
  \BibitemOpen
  \bibfield  {author} {\bibinfo {author} {\bibfnamefont {J.}~\bibnamefont
  {Guven}},\ }\href {https://doi.org/10.1088/0305-4470/37/28/L02} {\bibfield
  {journal} {\bibinfo  {journal} {J. Phys. A: Math. Gen.}\ }\textbf {\bibinfo
  {volume} {37}},\ \bibinfo {pages} {L313} (\bibinfo {year}
  {2004})}\BibitemShut {NoStop}%
\bibitem [{\citenamefont {Agrawal}\ and\ \citenamefont
  {Steigmann}(2008)}]{agrawal-steigmann-bmmb-2008}%
  \BibitemOpen
  \bibfield  {author} {\bibinfo {author} {\bibfnamefont {A.}~\bibnamefont
  {Agrawal}}\ and\ \bibinfo {author} {\bibfnamefont {D.~J.}\ \bibnamefont
  {Steigmann}},\ }\href {https://doi.org/10.1007/s10237-008-0143-0} {\bibfield
  {journal} {\bibinfo  {journal} {Biomech. Model. Mechan.}\ }\textbf {\bibinfo
  {volume} {8}},\ \bibinfo {pages} {371} (\bibinfo {year} {2008})}\BibitemShut
  {NoStop}%
\bibitem [{\citenamefont {Maitra}\ \emph {et~al.}(2014)\citenamefont {Maitra},
  \citenamefont {Srivastava}, \citenamefont {Rao},\ and\ \citenamefont
  {Ramaswamy}}]{ramaswamy-prl-2014}%
  \BibitemOpen
  \bibfield  {author} {\bibinfo {author} {\bibfnamefont {A.}~\bibnamefont
  {Maitra}}, \bibinfo {author} {\bibfnamefont {P.}~\bibnamefont {Srivastava}},
  \bibinfo {author} {\bibfnamefont {M.}~\bibnamefont {Rao}}, \ and\ \bibinfo
  {author} {\bibfnamefont {S.}~\bibnamefont {Ramaswamy}},\ }\href
  {https://doi.org/10.1103/PhysRevLett.112.258101} {\bibfield  {journal}
  {\bibinfo  {journal} {Phys. Rev. Lett.}\ }\textbf {\bibinfo {volume} {112}},\
  \bibinfo {pages} {258101} (\bibinfo {year} {2014})}\BibitemShut {NoStop}%
\bibitem [{\citenamefont {Al-Izzi}\ \emph {et~al.}(2018)\citenamefont
  {Al-Izzi}, \citenamefont {Rowlands}, \citenamefont {Sens},\ and\
  \citenamefont {Turner}}]{al-izzi-prl-2018}%
  \BibitemOpen
  \bibfield  {author} {\bibinfo {author} {\bibfnamefont {S.~C.}\ \bibnamefont
  {Al-Izzi}}, \bibinfo {author} {\bibfnamefont {G.}~\bibnamefont {Rowlands}},
  \bibinfo {author} {\bibfnamefont {P.}~\bibnamefont {Sens}}, \ and\ \bibinfo
  {author} {\bibfnamefont {M.~S.}\ \bibnamefont {Turner}},\ }\href
  {https://dx.doi.org/10.1103/PhysRevLett.120.138102} {\bibfield  {journal}
  {\bibinfo  {journal} {Phys. Rev. Lett.}\ }\textbf {\bibinfo {volume} {120}},\
  \bibinfo {pages} {138102} (\bibinfo {year} {2018})}\BibitemShut {NoStop}%
\bibitem [{\citenamefont {Bar-Ziv}\ and\ \citenamefont
  {Moses}(1994)}]{bar-ziv-prl-1994}%
  \BibitemOpen
  \bibfield  {author} {\bibinfo {author} {\bibfnamefont {R.}~\bibnamefont
  {Bar-Ziv}}\ and\ \bibinfo {author} {\bibfnamefont {E.}~\bibnamefont
  {Moses}},\ }\href {https://dx.doi.org/10.1103/PhysRevLett.73.1392} {\bibfield
   {journal} {\bibinfo  {journal} {Phys. Rev. Lett.}\ }\textbf {\bibinfo
  {volume} {73}},\ \bibinfo {pages} {1392} (\bibinfo {year}
  {1994})}\BibitemShut {NoStop}%
\bibitem [{\citenamefont {Goldstein}\ \emph {et~al.}(1996)\citenamefont
  {Goldstein}, \citenamefont {Nelson}, \citenamefont {Powers},\ and\
  \citenamefont {Seifert}}]{goldstein-jp-1996}%
  \BibitemOpen
  \bibfield  {author} {\bibinfo {author} {\bibfnamefont {R.~E.}\ \bibnamefont
  {Goldstein}}, \bibinfo {author} {\bibfnamefont {P.}~\bibnamefont {Nelson}},
  \bibinfo {author} {\bibfnamefont {T.}~\bibnamefont {Powers}}, \ and\ \bibinfo
  {author} {\bibfnamefont {U.}~\bibnamefont {Seifert}},\ }\href
  {https://doi.org/10.1051/jp2:1996210} {\bibfield  {journal} {\bibinfo
  {journal} {J. Phys. II}\ }\textbf {\bibinfo {volume} {6}},\ \bibinfo {pages}
  {767} (\bibinfo {year} {1996})}\BibitemShut {NoStop}%
\bibitem [{\citenamefont {Der\'enyi}\ \emph {et~al.}(2002)\citenamefont
  {Der\'enyi}, \citenamefont {J\"ulicher},\ and\ \citenamefont
  {Prost}}]{prost-prl-2002}%
  \BibitemOpen
  \bibfield  {author} {\bibinfo {author} {\bibfnamefont {I.}~\bibnamefont
  {Der\'enyi}}, \bibinfo {author} {\bibfnamefont {F.}~\bibnamefont
  {J\"ulicher}}, \ and\ \bibinfo {author} {\bibfnamefont {J.}~\bibnamefont
  {Prost}},\ }\href {https://doi.org/10.1103/PhysRevLett.88.238101} {\bibfield
  {journal} {\bibinfo  {journal} {Phys. Rev. Lett.}\ }\textbf {\bibinfo
  {volume} {88}},\ \bibinfo {pages} {238101} (\bibinfo {year}
  {2002})}\BibitemShut {NoStop}%
\bibitem [{\citenamefont {Kantsler}\ and\ \citenamefont
  {Steinberg}(2005)}]{kantsler-prl-2005}%
  \BibitemOpen
  \bibfield  {author} {\bibinfo {author} {\bibfnamefont {V.}~\bibnamefont
  {Kantsler}}\ and\ \bibinfo {author} {\bibfnamefont {V.}~\bibnamefont
  {Steinberg}},\ }\href {https://dx.doi.org/10.1103/PhysRevLett.95.258101}
  {\bibfield  {journal} {\bibinfo  {journal} {Phys. Rev. Lett.}\ }\textbf
  {\bibinfo {volume} {95}},\ \bibinfo {pages} {258101} (\bibinfo {year}
  {2005})}\BibitemShut {NoStop}%
\bibitem [{\citenamefont {Abkarian}\ and\ \citenamefont
  {Viallat}(2008)}]{viallat-sm-2008}%
  \BibitemOpen
  \bibfield  {author} {\bibinfo {author} {\bibfnamefont {M.}~\bibnamefont
  {Abkarian}}\ and\ \bibinfo {author} {\bibfnamefont {A.}~\bibnamefont
  {Viallat}},\ }\href {http://dx.doi.org/10.1039/B716612E} {\bibfield
  {journal} {\bibinfo  {journal} {Soft Matter}\ }\textbf {\bibinfo {volume}
  {4}},\ \bibinfo {pages} {653} (\bibinfo {year} {2008})}\BibitemShut {NoStop}%
\bibitem [{\citenamefont {Deschamps}\ \emph {et~al.}(2009)\citenamefont
  {Deschamps}, \citenamefont {Kantsler},\ and\ \citenamefont
  {Steinberg}}]{deschamps-prl-2009}%
  \BibitemOpen
  \bibfield  {author} {\bibinfo {author} {\bibfnamefont {J.}~\bibnamefont
  {Deschamps}}, \bibinfo {author} {\bibfnamefont {V.}~\bibnamefont {Kantsler}},
  \ and\ \bibinfo {author} {\bibfnamefont {V.}~\bibnamefont {Steinberg}},\
  }\href {https://dx.doi.org/10.1103/PhysRevLett.102.118105} {\bibfield
  {journal} {\bibinfo  {journal} {Phys. Rev. Lett.}\ }\textbf {\bibinfo
  {volume} {102}},\ \bibinfo {pages} {118105} (\bibinfo {year}
  {2009})}\BibitemShut {NoStop}%
\bibitem [{\citenamefont {Keller}\ and\ \citenamefont
  {Skalak}(1982)}]{keller-skalak-1982}%
  \BibitemOpen
  \bibfield  {author} {\bibinfo {author} {\bibfnamefont {S.~R.}\ \bibnamefont
  {Keller}}\ and\ \bibinfo {author} {\bibfnamefont {R.}~\bibnamefont
  {Skalak}},\ }\href {https://dx.doi.org/10.1017/S0022112082002651} {\bibfield
  {journal} {\bibinfo  {journal} {J. Fluid Mech.}\ }\textbf {\bibinfo {volume}
  {120}},\ \bibinfo {pages} {27} (\bibinfo {year} {1982})}\BibitemShut
  {NoStop}%
\bibitem [{\citenamefont {Kraus}\ \emph {et~al.}(1996)\citenamefont {Kraus},
  \citenamefont {Wintz}, \citenamefont {Seifert},\ and\ \citenamefont
  {Lipowsky}}]{kraus-prl-1996}%
  \BibitemOpen
  \bibfield  {author} {\bibinfo {author} {\bibfnamefont {M.}~\bibnamefont
  {Kraus}}, \bibinfo {author} {\bibfnamefont {W.}~\bibnamefont {Wintz}},
  \bibinfo {author} {\bibfnamefont {U.}~\bibnamefont {Seifert}}, \ and\
  \bibinfo {author} {\bibfnamefont {R.}~\bibnamefont {Lipowsky}},\ }\href
  {https://doi.org/10.1103/PhysRevLett.77.3685} {\bibfield  {journal} {\bibinfo
   {journal} {Phys. Rev. Lett.}\ }\textbf {\bibinfo {volume} {77}},\ \bibinfo
  {pages} {3685} (\bibinfo {year} {1996})}\BibitemShut {NoStop}%
\bibitem [{\citenamefont {Seifert}(1999)}]{seifert-epjb-1999}%
  \BibitemOpen
  \bibfield  {author} {\bibinfo {author} {\bibfnamefont {U.}~\bibnamefont
  {Seifert}},\ }\href {https://doi.org/10.1007/s100510050706} {\bibfield
  {journal} {\bibinfo  {journal} {Eur. Phys. J. B}\ }\textbf {\bibinfo {volume}
  {8}},\ \bibinfo {pages} {405} (\bibinfo {year} {1999})}\BibitemShut {NoStop}%
\bibitem [{\citenamefont {Beaucourt}\ \emph {et~al.}(2004)\citenamefont
  {Beaucourt}, \citenamefont {Rioual}, \citenamefont {S\'{e}on}, \citenamefont
  {Biben},\ and\ \citenamefont {Misbah}}]{beaucourt-pre-2004}%
  \BibitemOpen
  \bibfield  {author} {\bibinfo {author} {\bibfnamefont {J.}~\bibnamefont
  {Beaucourt}}, \bibinfo {author} {\bibfnamefont {F.}~\bibnamefont {Rioual}},
  \bibinfo {author} {\bibfnamefont {T.}~\bibnamefont {S\'{e}on}}, \bibinfo
  {author} {\bibfnamefont {T.}~\bibnamefont {Biben}}, \ and\ \bibinfo {author}
  {\bibfnamefont {C.}~\bibnamefont {Misbah}},\ }\href
  {https://dx.doi.org/10.1103/PhysRevE.69.011906} {\bibfield  {journal}
  {\bibinfo  {journal} {Phys. Rev. E}\ }\textbf {\bibinfo {volume} {69}},\
  \bibinfo {pages} {011906} (\bibinfo {year} {2004})}\BibitemShut {NoStop}%
\bibitem [{\citenamefont {Noguchi}\ and\ \citenamefont
  {Gompper}(2007)}]{noguchi-prl-2007}%
  \BibitemOpen
  \bibfield  {author} {\bibinfo {author} {\bibfnamefont {H.}~\bibnamefont
  {Noguchi}}\ and\ \bibinfo {author} {\bibfnamefont {G.}~\bibnamefont
  {Gompper}},\ }\href {https://dx.doi.org/10.1103/PhysRevLett.98.128103}
  {\bibfield  {journal} {\bibinfo  {journal} {Phys. Rev. Lett.}\ }\textbf
  {\bibinfo {volume} {98}},\ \bibinfo {pages} {128103} (\bibinfo {year}
  {2007})}\BibitemShut {NoStop}%
\bibitem [{\citenamefont {Vlahovska}\ and\ \citenamefont
  {Gracia}(2007)}]{vlahovska-pre-2007}%
  \BibitemOpen
  \bibfield  {author} {\bibinfo {author} {\bibfnamefont {P.~M.}\ \bibnamefont
  {Vlahovska}}\ and\ \bibinfo {author} {\bibfnamefont {R.~S.}\ \bibnamefont
  {Gracia}},\ }\href {https://dx.doi.org/10.1103/PhysRevE.75.016313} {\bibfield
   {journal} {\bibinfo  {journal} {Phys. Rev. E}\ }\textbf {\bibinfo {volume}
  {75}},\ \bibinfo {pages} {016313} (\bibinfo {year} {2007})}\BibitemShut
  {NoStop}%
\bibitem [{\citenamefont {Me\ss{}linger}\ \emph {et~al.}(2009)\citenamefont
  {Me\ss{}linger}, \citenamefont {Schmidt}, \citenamefont {Noguchi},\ and\
  \citenamefont {Gompper}}]{messlinger-pre-2009}%
  \BibitemOpen
  \bibfield  {author} {\bibinfo {author} {\bibfnamefont {S.}~\bibnamefont
  {Me\ss{}linger}}, \bibinfo {author} {\bibfnamefont {B.}~\bibnamefont
  {Schmidt}}, \bibinfo {author} {\bibfnamefont {H.}~\bibnamefont {Noguchi}}, \
  and\ \bibinfo {author} {\bibfnamefont {G.}~\bibnamefont {Gompper}},\ }\href
  {\doibase https://dx.doi.org/10.1103/PhysRevE.80.011901} {\bibfield
  {journal} {\bibinfo  {journal} {Phys. Rev. E}\ }\textbf {\bibinfo {volume}
  {80}},\ \bibinfo {pages} {011901} (\bibinfo {year} {2009})}\BibitemShut
  {NoStop}%
\bibitem [{\citenamefont {Zhao}\ and\ \citenamefont
  {Shaqfeh}(2011)}]{zhao-shaqfeh-jfm-2011}%
  \BibitemOpen
  \bibfield  {author} {\bibinfo {author} {\bibfnamefont {H.}~\bibnamefont
  {Zhao}}\ and\ \bibinfo {author} {\bibfnamefont {E.~S.~G.}\ \bibnamefont
  {Shaqfeh}},\ }\href {https://dx.doi.org/10.1017/S0022112011000115} {\bibfield
   {journal} {\bibinfo  {journal} {J. Fluid Mech.}\ }\textbf {\bibinfo {volume}
  {674}},\ \bibinfo {pages} {578} (\bibinfo {year} {2011})}\BibitemShut
  {NoStop}%
\bibitem [{\citenamefont {Zhao}\ \emph {et~al.}(2011)\citenamefont {Zhao},
  \citenamefont {Spann},\ and\ \citenamefont {Shaqfeh}}]{zhao-shaqfeh-pf-2011}%
  \BibitemOpen
  \bibfield  {author} {\bibinfo {author} {\bibfnamefont {H.}~\bibnamefont
  {Zhao}}, \bibinfo {author} {\bibfnamefont {A.~P.}\ \bibnamefont {Spann}}, \
  and\ \bibinfo {author} {\bibfnamefont {E.~S.~G.}\ \bibnamefont {Shaqfeh}},\
  }\href {https://doi.org/10.1063/1.3669440} {\bibfield  {journal} {\bibinfo
  {journal} {Phys. Fluids}\ }\textbf {\bibinfo {volume} {23}},\ \bibinfo
  {pages} {121901} (\bibinfo {year} {2011})}\BibitemShut {NoStop}%
\bibitem [{\citenamefont {Arroyo}\ and\ \citenamefont
  {DeSimone}(2009)}]{arroyo-pre-2009}%
  \BibitemOpen
  \bibfield  {author} {\bibinfo {author} {\bibfnamefont {M.}~\bibnamefont
  {Arroyo}}\ and\ \bibinfo {author} {\bibfnamefont {A.}~\bibnamefont
  {DeSimone}},\ }\href {https://doi.org/10.1103/PhysRevE.79.031915} {\bibfield
  {journal} {\bibinfo  {journal} {Phys. Rev. E}\ }\textbf {\bibinfo {volume}
  {79}},\ \bibinfo {pages} {31915} (\bibinfo {year} {2009})}\BibitemShut
  {NoStop}%
\bibitem [{\citenamefont {Powers}(2010)}]{powers-rmp-2010}%
  \BibitemOpen
  \bibfield  {author} {\bibinfo {author} {\bibfnamefont {T.~R.}\ \bibnamefont
  {Powers}},\ }\href {https://dx.doi.org/10.1103/RevModPhys.82.1607} {\bibfield
   {journal} {\bibinfo  {journal} {Rev. Mod. Phys.}\ }\textbf {\bibinfo
  {volume} {82}},\ \bibinfo {pages} {1607} (\bibinfo {year}
  {2010})}\BibitemShut {NoStop}%
\bibitem [{\citenamefont {Rangamani}\ \emph {et~al.}(2012)\citenamefont
  {Rangamani}, \citenamefont {Agrawal}, \citenamefont {Mandadapu},
  \citenamefont {Oster},\ and\ \citenamefont {Steigmann}}]{kranthi-bmmb-2012}%
  \BibitemOpen
  \bibfield  {author} {\bibinfo {author} {\bibfnamefont {P.}~\bibnamefont
  {Rangamani}}, \bibinfo {author} {\bibfnamefont {A.}~\bibnamefont {Agrawal}},
  \bibinfo {author} {\bibfnamefont {K.~K.}\ \bibnamefont {Mandadapu}}, \bibinfo
  {author} {\bibfnamefont {G.}~\bibnamefont {Oster}}, \ and\ \bibinfo {author}
  {\bibfnamefont {D.~J.}\ \bibnamefont {Steigmann}},\ }\href
  {https://doi.org/10.1007/s10237-012-0447-y} {\bibfield  {journal} {\bibinfo
  {journal} {Biomech. Model. Mechan.}\ }\textbf {\bibinfo {volume} {12}},\
  \bibinfo {pages} {833} (\bibinfo {year} {2012})}\BibitemShut {NoStop}%
\bibitem [{\citenamefont {Sahu}\ \emph {et~al.}(2017)\citenamefont {Sahu},
  \citenamefont {Sauer},\ and\ \citenamefont
  {Mandadapu}}]{sahu-mandadapu-pre-2017}%
  \BibitemOpen
  \bibfield  {author} {\bibinfo {author} {\bibfnamefont {A.}~\bibnamefont
  {Sahu}}, \bibinfo {author} {\bibfnamefont {R.~A.}\ \bibnamefont {Sauer}}, \
  and\ \bibinfo {author} {\bibfnamefont {K.~K.}\ \bibnamefont {Mandadapu}},\
  }\href {https://doi.org/10.1103/PhysRevE.96.042409} {\bibfield  {journal}
  {\bibinfo  {journal} {Phys. Rev. E}\ }\textbf {\bibinfo {volume} {96}},\
  \bibinfo {pages} {042409} (\bibinfo {year} {2017})},\ \Eprint
  {http://arxiv.org/abs/1701.06495} {arXiv:1701.06495} \BibitemShut {NoStop}%
\bibitem [{\citenamefont {Lidmar}\ \emph {et~al.}(2003)\citenamefont {Lidmar},
  \citenamefont {Mirny},\ and\ \citenamefont {Nelson}}]{nelson-pre-2003}%
  \BibitemOpen
  \bibfield  {author} {\bibinfo {author} {\bibfnamefont {J.}~\bibnamefont
  {Lidmar}}, \bibinfo {author} {\bibfnamefont {L.}~\bibnamefont {Mirny}}, \
  and\ \bibinfo {author} {\bibfnamefont {D.~R.}\ \bibnamefont {Nelson}},\
  }\href {https://dx.doi.org/10.1103/PhysRevE.68.051910} {\bibfield  {journal}
  {\bibinfo  {journal} {Phys. Rev. E}\ }\textbf {\bibinfo {volume} {68}},\
  \bibinfo {pages} {051910} (\bibinfo {year} {2003})}\BibitemShut {NoStop}%
\bibitem [{\citenamefont {Scriven}(1960)}]{scriven-1960}%
  \BibitemOpen
  \bibfield  {author} {\bibinfo {author} {\bibfnamefont {L.~E.}\ \bibnamefont
  {Scriven}},\ }\href {https://doi.org/10.1016/0009-2509(60)87003-0} {\bibfield
   {journal} {\bibinfo  {journal} {Chem. Eng. Sci.}\ }\textbf {\bibinfo
  {volume} {12}},\ \bibinfo {pages} {98} (\bibinfo {year} {1960})}\BibitemShut
  {NoStop}%
\bibitem [{\citenamefont {Love}(1927)}]{love}%
  \BibitemOpen
  \bibfield  {author} {\bibinfo {author} {\bibfnamefont {A.~E.~H.}\
  \bibnamefont {Love}},\ }\href@noop {} {\emph {\bibinfo {title} {A Treatise on
  the Mathematical Theory of Elasticity}}}\ (\bibinfo  {publisher} {Cambridge
  University Press},\ \bibinfo {address} {Cambridge},\ \bibinfo {year}
  {1927})\BibitemShut {NoStop}%
\bibitem [{sup()}]{supplemental}%
  \BibitemOpen
  \href {https://amaresh-sahu.github.io/papers/sahu-mandadapu-geo-dyn-SM.pdf}
  {\bibfield  {journal} {\bibinfo  {journal} {See the Supplemental Material,}\
  }}\bibinfo {note} {{in} which the equations of motion are determined and
  non-dimensionalized about planar, spherical, and cylindrical shapes. The
  Supplemental Material additionally cites Refs.\ \cite{kreyszig, willmore,
  naghdi-1973-theory, parkkila-langmuir-2018}.}\BibitemShut {Stop}%
\bibitem [{\citenamefont {Nichol}\ and\ \citenamefont
  {Hutter}(1996)}]{nichol-jp-1996}%
  \BibitemOpen
  \bibfield  {author} {\bibinfo {author} {\bibfnamefont {J.~A.}\ \bibnamefont
  {Nichol}}\ and\ \bibinfo {author} {\bibfnamefont {O.~F.}\ \bibnamefont
  {Hutter}},\ }\href {https://doi.org/10.1113/jphysiol.1996.sp021374}
  {\bibfield  {journal} {\bibinfo  {journal} {J. Physiol.}\ }\textbf {\bibinfo
  {volume} {493}},\ \bibinfo {pages} {187} (\bibinfo {year}
  {1996})}\BibitemShut {NoStop}%
\bibitem [{\citenamefont {Edwards}\ \emph {et~al.}(1991)\citenamefont
  {Edwards}, \citenamefont {Brenner},\ and\ \citenamefont
  {Wasan}}]{edwards-brenner}%
  \BibitemOpen
  \bibfield  {author} {\bibinfo {author} {\bibfnamefont {D.~A.}\ \bibnamefont
  {Edwards}}, \bibinfo {author} {\bibfnamefont {H.}~\bibnamefont {Brenner}}, \
  and\ \bibinfo {author} {\bibfnamefont {D.}~\bibnamefont {Wasan}},\ }\href
  {https://doi.org/10.1016/C2009-0-26916-9} {\emph {\bibinfo {title}
  {Interfacial Transport Processes and Rheology}}},\ Butterworth-Heinemann
  Series in Chemical Engineering\ (\bibinfo  {publisher}
  {Butterworth-Heinemann},\ \bibinfo {address} {Boston},\ \bibinfo {year}
  {1991})\BibitemShut {NoStop}%
\bibitem [{\citenamefont {Canham}(1970)}]{canham-jtb-1970}%
  \BibitemOpen
  \bibfield  {author} {\bibinfo {author} {\bibfnamefont {P.~B.}\ \bibnamefont
  {Canham}},\ }\href {https://doi.org/10.1016/S0022-5193(70)80032-7} {\bibfield
   {journal} {\bibinfo  {journal} {J. Theor. Biol.}\ }\textbf {\bibinfo
  {volume} {26}},\ \bibinfo {pages} {61} (\bibinfo {year} {1970})}\BibitemShut
  {NoStop}%
\bibitem [{\citenamefont {Helfrich}(1973)}]{helfrich-1973}%
  \BibitemOpen
  \bibfield  {author} {\bibinfo {author} {\bibfnamefont {W.}~\bibnamefont
  {Helfrich}},\ }\href {https://doi.org/10.1515/znc-1973-11-1209} {\bibfield
  {journal} {\bibinfo  {journal} {Z. Naturforsch. C}\ }\textbf {\bibinfo
  {volume} {28}},\ \bibinfo {pages} {693} (\bibinfo {year} {1973})}\BibitemShut
  {NoStop}%
\bibitem [{\citenamefont {Evans}(1974)}]{evans-bpj-1974}%
  \BibitemOpen
  \bibfield  {author} {\bibinfo {author} {\bibfnamefont {E.~A.}\ \bibnamefont
  {Evans}},\ }\href {https://doi.org/10.1016/S0006-3495(74)85959-X} {\bibfield
  {journal} {\bibinfo  {journal} {Biophys. J.}\ }\textbf {\bibinfo {volume}
  {14}},\ \bibinfo {pages} {923} (\bibinfo {year} {1974})}\BibitemShut
  {NoStop}%
\bibitem [{\citenamefont {Watanabe}\ \emph
  {et~al.}(2013{\natexlab{b}})\citenamefont {Watanabe}, \citenamefont {Liu},
  \citenamefont {Davis}, \citenamefont {Hollopeter}, \citenamefont {Thomas},
  \citenamefont {Jorgensen},\ and\ \citenamefont
  {Jorgensen}}]{watanabe-elife-2013}%
  \BibitemOpen
  \bibfield  {author} {\bibinfo {author} {\bibfnamefont {S.}~\bibnamefont
  {Watanabe}}, \bibinfo {author} {\bibfnamefont {Q.}~\bibnamefont {Liu}},
  \bibinfo {author} {\bibfnamefont {M.~W.}\ \bibnamefont {Davis}}, \bibinfo
  {author} {\bibfnamefont {G.}~\bibnamefont {Hollopeter}}, \bibinfo {author}
  {\bibfnamefont {N.}~\bibnamefont {Thomas}}, \bibinfo {author} {\bibfnamefont
  {N.~B.}\ \bibnamefont {Jorgensen}}, \ and\ \bibinfo {author} {\bibfnamefont
  {E.~M.}\ \bibnamefont {Jorgensen}},\ }\href
  {https://doi.org/10.7554/eLife.00723} {\bibfield  {journal} {\bibinfo
  {journal} {eLife}\ }\textbf {\bibinfo {volume} {2}},\ \bibinfo {pages}
  {e00723} (\bibinfo {year} {2013}{\natexlab{b}})}\BibitemShut {NoStop}%
\bibitem [{\citenamefont {Cocucci}\ \emph {et~al.}(2012)\citenamefont
  {Cocucci}, \citenamefont {Aguet}, \citenamefont {Boulant},\ and\
  \citenamefont {Kirchhausen}}]{cocucci-cell-2012}%
  \BibitemOpen
  \bibfield  {author} {\bibinfo {author} {\bibfnamefont {E.}~\bibnamefont
  {Cocucci}}, \bibinfo {author} {\bibfnamefont {F.}~\bibnamefont {Aguet}},
  \bibinfo {author} {\bibfnamefont {S.}~\bibnamefont {Boulant}}, \ and\
  \bibinfo {author} {\bibfnamefont {T.}~\bibnamefont {Kirchhausen}},\ }\href
  {https://doi.org/10.1016/j.cell.2012.05.047} {\bibfield  {journal} {\bibinfo
  {journal} {Cell}\ }\textbf {\bibinfo {volume} {150}},\ \bibinfo {pages} {495}
  (\bibinfo {year} {2012})}\BibitemShut {NoStop}%
\bibitem [{\citenamefont {de~Haas}\ \emph {et~al.}(1997)\citenamefont
  {de~Haas}, \citenamefont {Blom}, \citenamefont {van~den Ende}, \citenamefont
  {Duits},\ and\ \citenamefont {Mellema}}]{mellema-pre-1997}%
  \BibitemOpen
  \bibfield  {author} {\bibinfo {author} {\bibfnamefont {K.~H.}\ \bibnamefont
  {de~Haas}}, \bibinfo {author} {\bibfnamefont {C.}~\bibnamefont {Blom}},
  \bibinfo {author} {\bibfnamefont {D.}~\bibnamefont {van~den Ende}}, \bibinfo
  {author} {\bibfnamefont {M.~H.~G.}\ \bibnamefont {Duits}}, \ and\ \bibinfo
  {author} {\bibfnamefont {J.}~\bibnamefont {Mellema}},\ }\href
  {https://dx.doi.org/10.1103/PhysRevE.56.7132} {\bibfield  {journal} {\bibinfo
   {journal} {Phys. Rev. E}\ }\textbf {\bibinfo {volume} {56}},\ \bibinfo
  {pages} {7132} (\bibinfo {year} {1997})}\BibitemShut {NoStop}%
\bibitem [{\citenamefont {Lipowsky}\ \emph {et~al.}(1980)\citenamefont
  {Lipowsky}, \citenamefont {Usami},\ and\ \citenamefont
  {Chien}}]{lipowsky-mr-1980}%
  \BibitemOpen
  \bibfield  {author} {\bibinfo {author} {\bibfnamefont {H.~H.}\ \bibnamefont
  {Lipowsky}}, \bibinfo {author} {\bibfnamefont {S.}~\bibnamefont {Usami}}, \
  and\ \bibinfo {author} {\bibfnamefont {S.}~\bibnamefont {Chien}},\ }\href
  {https://doi.org/10.1016/0026-2862(80)90050-3} {\bibfield  {journal}
  {\bibinfo  {journal} {Microvasc. Res.}\ }\textbf {\bibinfo {volume} {19}},\
  \bibinfo {pages} {297} (\bibinfo {year} {1980})}\BibitemShut {NoStop}%
\bibitem [{\citenamefont {Kolaczkowska}\ and\ \citenamefont
  {Kubes}(2013)}]{kolaczkowska-nri-2013}%
  \BibitemOpen
  \bibfield  {author} {\bibinfo {author} {\bibfnamefont {E.}~\bibnamefont
  {Kolaczkowska}}\ and\ \bibinfo {author} {\bibfnamefont {P.}~\bibnamefont
  {Kubes}},\ }\href {https://doi.org/10.1038/nri3399} {\bibfield  {journal}
  {\bibinfo  {journal} {Nat. Rev. Immunol.}\ }\textbf {\bibinfo {volume} {13}}
  (\bibinfo {year} {2013})}\BibitemShut {NoStop}%
\bibitem [{\citenamefont {Ota}\ \emph {et~al.}(2009)\citenamefont {Ota},
  \citenamefont {Yoshizawa},\ and\ \citenamefont {Takeuchi}}]{ota-ac-2009}%
  \BibitemOpen
  \bibfield  {author} {\bibinfo {author} {\bibfnamefont {S.}~\bibnamefont
  {Ota}}, \bibinfo {author} {\bibfnamefont {S.}~\bibnamefont {Yoshizawa}}, \
  and\ \bibinfo {author} {\bibfnamefont {S.}~\bibnamefont {Takeuchi}},\ }\href
  {https://dx.doi.org/10.1002/anie.200902182} {\bibfield  {journal} {\bibinfo
  {journal} {Angew. Chem. Int. Edit.}\ }\textbf {\bibinfo {volume} {48}},\
  \bibinfo {pages} {6533} (\bibinfo {year} {2009})}\BibitemShut {NoStop}%
\bibitem [{\citenamefont {Mader}\ \emph {et~al.}(2006)\citenamefont {Mader},
  \citenamefont {Vitkova}, \citenamefont {Abkarian}, \citenamefont {Viallat},\
  and\ \citenamefont {Podgorski}}]{mader-epje-2006}%
  \BibitemOpen
  \bibfield  {author} {\bibinfo {author} {\bibfnamefont {M.-A.}\ \bibnamefont
  {Mader}}, \bibinfo {author} {\bibfnamefont {V.}~\bibnamefont {Vitkova}},
  \bibinfo {author} {\bibfnamefont {M.}~\bibnamefont {Abkarian}}, \bibinfo
  {author} {\bibfnamefont {A.}~\bibnamefont {Viallat}}, \ and\ \bibinfo
  {author} {\bibfnamefont {T.}~\bibnamefont {Podgorski}},\ }\href
  {https://doi.org/10.1140/epje/i2005-10058-x} {\bibfield  {journal} {\bibinfo
  {journal} {Eur. Phys. J. E}\ }\textbf {\bibinfo {volume} {19}},\ \bibinfo
  {pages} {389} (\bibinfo {year} {2006})}\BibitemShut {NoStop}%
\bibitem [{\citenamefont {F{\"o}rster}\ \emph {et~al.}(2005)\citenamefont
  {F{\"o}rster}, \citenamefont {Medalia}, \citenamefont {Zauberman},
  \citenamefont {Baumeister},\ and\ \citenamefont {Fass}}]{forster-pnas-2005}%
  \BibitemOpen
  \bibfield  {author} {\bibinfo {author} {\bibfnamefont {F.}~\bibnamefont
  {F{\"o}rster}}, \bibinfo {author} {\bibfnamefont {O.}~\bibnamefont
  {Medalia}}, \bibinfo {author} {\bibfnamefont {N.}~\bibnamefont {Zauberman}},
  \bibinfo {author} {\bibfnamefont {W.}~\bibnamefont {Baumeister}}, \ and\
  \bibinfo {author} {\bibfnamefont {D.}~\bibnamefont {Fass}},\ }\href
  {https://dx.doi.org/10.1073/pnas.0409178102} {\bibfield  {journal} {\bibinfo
  {journal} {Proc. Natl. Acad. Sci. U.S.A.}\ }\textbf {\bibinfo {volume}
  {102}},\ \bibinfo {pages} {4729} (\bibinfo {year} {2005})}\BibitemShut
  {NoStop}%
\bibitem [{\citenamefont {Shi}\ \emph {et~al.}(2018)\citenamefont {Shi},
  \citenamefont {Graber}, \citenamefont {Baumgart}, \citenamefont {Stone},\
  and\ \citenamefont {Cohen}}]{shi-cell-2018}%
  \BibitemOpen
  \bibfield  {author} {\bibinfo {author} {\bibfnamefont {Z.}~\bibnamefont
  {Shi}}, \bibinfo {author} {\bibfnamefont {Z.~T.}\ \bibnamefont {Graber}},
  \bibinfo {author} {\bibfnamefont {T.}~\bibnamefont {Baumgart}}, \bibinfo
  {author} {\bibfnamefont {H.~A.}\ \bibnamefont {Stone}}, \ and\ \bibinfo
  {author} {\bibfnamefont {A.~E.}\ \bibnamefont {Cohen}},\ }\href
  {https://doi.org/10.1016/j.cell.2018.09.054} {\bibfield  {journal} {\bibinfo
  {journal} {Cell}\ }\textbf {\bibinfo {volume} {175}},\ \bibinfo {pages}
  {1769} (\bibinfo {year} {2018})}\BibitemShut {NoStop}%
\bibitem [{\citenamefont {Honerkamp-Smith}\ \emph {et~al.}(2013)\citenamefont
  {Honerkamp-Smith}, \citenamefont {Woodhouse}, \citenamefont {Kantsler},\ and\
  \citenamefont {Goldstein}}]{honerkamp-prl-2013}%
  \BibitemOpen
  \bibfield  {author} {\bibinfo {author} {\bibfnamefont {A.~R.}\ \bibnamefont
  {Honerkamp-Smith}}, \bibinfo {author} {\bibfnamefont {F.~G.}\ \bibnamefont
  {Woodhouse}}, \bibinfo {author} {\bibfnamefont {V.}~\bibnamefont {Kantsler}},
  \ and\ \bibinfo {author} {\bibfnamefont {R.~E.}\ \bibnamefont {Goldstein}},\
  }\href {https://dx.doi.org/10.1103/PhysRevLett.111.038103} {\bibfield
  {journal} {\bibinfo  {journal} {Phys. Rev. Lett.}\ }\textbf {\bibinfo
  {volume} {111}},\ \bibinfo {pages} {038103} (\bibinfo {year}
  {2013})}\BibitemShut {NoStop}%
\bibitem [{\citenamefont {Boedec}\ \emph {et~al.}(2014)\citenamefont {Boedec},
  \citenamefont {Jaeger},\ and\ \citenamefont {Leonetti}}]{boedec-jfm-2014}%
  \BibitemOpen
  \bibfield  {author} {\bibinfo {author} {\bibfnamefont {G.}~\bibnamefont
  {Boedec}}, \bibinfo {author} {\bibfnamefont {M.}~\bibnamefont {Jaeger}}, \
  and\ \bibinfo {author} {\bibfnamefont {M.}~\bibnamefont {Leonetti}},\ }\href
  {https://dx.doi.org/10.1017/jfm.2014.34} {\bibfield  {journal} {\bibinfo
  {journal} {J. Fluid Mech.}\ }\textbf {\bibinfo {volume} {743}},\ \bibinfo
  {pages} {262} (\bibinfo {year} {2014})}\BibitemShut {NoStop}%
\bibitem [{\citenamefont {Narsimhan}\ \emph {et~al.}(2015)\citenamefont
  {Narsimhan}, \citenamefont {Spann},\ and\ \citenamefont
  {Shaqfeh}}]{narsimhan-shaqfeh-jfm-2015}%
  \BibitemOpen
  \bibfield  {author} {\bibinfo {author} {\bibfnamefont {V.}~\bibnamefont
  {Narsimhan}}, \bibinfo {author} {\bibfnamefont {A.}~\bibnamefont {Spann}}, \
  and\ \bibinfo {author} {\bibfnamefont {E.}~\bibnamefont {Shaqfeh}},\ }\href
  {https://dx.doi.org/10.1017/jfm.2015.345} {\bibfield  {journal} {\bibinfo
  {journal} {J. Fluid Mech.}\ }\textbf {\bibinfo {volume} {777}},\ \bibinfo
  {pages} {1} (\bibinfo {year} {2015})}\BibitemShut {NoStop}%
\bibitem [{\citenamefont {Monge}(1807)}]{monge}%
  \BibitemOpen
  \bibfield  {author} {\bibinfo {author} {\bibfnamefont {G.}~\bibnamefont
  {Monge}},\ }\href@noop {} {\emph {\bibinfo {title} {Application de l'analyse
  \`{a} la g\'{e}om\'{e}trie}}}\ (\bibinfo  {publisher} {Bernard},\ \bibinfo
  {address} {Paris},\ \bibinfo {year} {1807})\BibitemShut {NoStop}%
\bibitem [{\citenamefont {Seifert}\ and\ \citenamefont
  {Langer}(1993)}]{seifert-epl-1993}%
  \BibitemOpen
  \bibfield  {author} {\bibinfo {author} {\bibfnamefont {U.}~\bibnamefont
  {Seifert}}\ and\ \bibinfo {author} {\bibfnamefont {S.~A.}\ \bibnamefont
  {Langer}},\ }\href {https://doi.org/10.1209/0295-5075/23/1/012} {\bibfield
  {journal} {\bibinfo  {journal} {Europhys. Lett.}\ }\textbf {\bibinfo {volume}
  {23}},\ \bibinfo {pages} {71} (\bibinfo {year} {1993})}\BibitemShut {NoStop}%
\bibitem [{\citenamefont {P{\'e}cr{\'e}aux}\ \emph {et~al.}(2004)\citenamefont
  {P{\'e}cr{\'e}aux}, \citenamefont {D{\"o}bereiner}, \citenamefont {Prost},
  \citenamefont {Joanny},\ and\ \citenamefont
  {Bassereau}}]{pecreaux-bassereau-epje-2004}%
  \BibitemOpen
  \bibfield  {author} {\bibinfo {author} {\bibfnamefont {J.}~\bibnamefont
  {P{\'e}cr{\'e}aux}}, \bibinfo {author} {\bibfnamefont {H.-G.}\ \bibnamefont
  {D{\"o}bereiner}}, \bibinfo {author} {\bibfnamefont {J.}~\bibnamefont
  {Prost}}, \bibinfo {author} {\bibfnamefont {J.-F.}\ \bibnamefont {Joanny}}, \
  and\ \bibinfo {author} {\bibfnamefont {P.}~\bibnamefont {Bassereau}},\ }\href
  {https://doi.org/10.1140/epje/i2004-10001-9} {\bibfield  {journal} {\bibinfo
  {journal} {Eur. Phys. J. E}\ }\textbf {\bibinfo {volume} {13}},\ \bibinfo
  {pages} {277} (\bibinfo {year} {2004})}\BibitemShut {NoStop}%
\bibitem [{\citenamefont {Dai}\ \emph {et~al.}(1998)\citenamefont {Dai},
  \citenamefont {Sheetz}, \citenamefont {Wan},\ and\ \citenamefont
  {Morris}}]{dai-jn-1998}%
  \BibitemOpen
  \bibfield  {author} {\bibinfo {author} {\bibfnamefont {J.}~\bibnamefont
  {Dai}}, \bibinfo {author} {\bibfnamefont {M.~P.}\ \bibnamefont {Sheetz}},
  \bibinfo {author} {\bibfnamefont {X.}~\bibnamefont {Wan}}, \ and\ \bibinfo
  {author} {\bibfnamefont {C.~E.}\ \bibnamefont {Morris}},\ }\href
  {https://doi.org/10.1523/JNEUROSCI.18-17-06681.1998} {\bibfield  {journal}
  {\bibinfo  {journal} {J. Neurosci.}\ }\textbf {\bibinfo {volume} {18}},\
  \bibinfo {pages} {6681} (\bibinfo {year} {1998})}\BibitemShut {NoStop}%
\bibitem [{\citenamefont {Higgins}\ and\ \citenamefont
  {McMahon}(2002)}]{mcmahon-cell-2002}%
  \BibitemOpen
  \bibfield  {author} {\bibinfo {author} {\bibfnamefont {M.~K.}\ \bibnamefont
  {Higgins}}\ and\ \bibinfo {author} {\bibfnamefont {H.~T.}\ \bibnamefont
  {McMahon}},\ }\href {https://doi.org/10.1016/S0968-0004(02)02089-3}
  {\bibfield  {journal} {\bibinfo  {journal} {Trends Biochem. Sci.}\ }\textbf
  {\bibinfo {volume} {27}},\ \bibinfo {pages} {257} (\bibinfo {year}
  {2002})}\BibitemShut {NoStop}%
\bibitem [{\citenamefont {Zhang}\ and\ \citenamefont
  {Jackson}(2010)}]{zhang-bpj-2010}%
  \BibitemOpen
  \bibfield  {author} {\bibinfo {author} {\bibfnamefont {Z.}~\bibnamefont
  {Zhang}}\ and\ \bibinfo {author} {\bibfnamefont {M.~B.}\ \bibnamefont
  {Jackson}},\ }\href {https://doi.org/10.1016/j.bpj.2010.02.043} {\bibfield
  {journal} {\bibinfo  {journal} {Biophys. J.}\ }\textbf {\bibinfo {volume}
  {98}},\ \bibinfo {pages} {2524} (\bibinfo {year} {2010})}\BibitemShut
  {NoStop}%
\bibitem [{\citenamefont {Allen}\ and\ \citenamefont
  {Aderem}(1996)}]{allen-coi-1996}%
  \BibitemOpen
  \bibfield  {author} {\bibinfo {author} {\bibfnamefont {L.-A.~H.}\
  \bibnamefont {Allen}}\ and\ \bibinfo {author} {\bibfnamefont
  {A.}~\bibnamefont {Aderem}},\ }\href
  {https://doi.org/10.1016/S0952-7915(96)80102-6} {\bibfield  {journal}
  {\bibinfo  {journal} {Curr. Opin. Immunol.}\ }\textbf {\bibinfo {volume}
  {8}},\ \bibinfo {pages} {36} (\bibinfo {year} {1996})}\BibitemShut {NoStop}%
\bibitem [{\citenamefont {Lee}\ and\ \citenamefont
  {Chen}(1988)}]{lee-cell-1988}%
  \BibitemOpen
  \bibfield  {author} {\bibinfo {author} {\bibfnamefont {C.}~\bibnamefont
  {Lee}}\ and\ \bibinfo {author} {\bibfnamefont {L.-B.}\ \bibnamefont {Chen}},\
  }\href {https://doi.org/10.1016/0092-8674(88)90177-8} {\bibfield  {journal}
  {\bibinfo  {journal} {Cell}\ }\textbf {\bibinfo {volume} {54}},\ \bibinfo
  {pages} {37} (\bibinfo {year} {1988})}\BibitemShut {NoStop}%
\bibitem [{\citenamefont {Dahl}\ \emph {et~al.}(2016)\citenamefont {Dahl},
  \citenamefont {Narsimhan}, \citenamefont {Gouveia}, \citenamefont {Kumar},
  \citenamefont {Shaqfeh},\ and\ \citenamefont {Muller}}]{dahl-sm-2016}%
  \BibitemOpen
  \bibfield  {author} {\bibinfo {author} {\bibfnamefont {J.~B.}\ \bibnamefont
  {Dahl}}, \bibinfo {author} {\bibfnamefont {V.}~\bibnamefont {Narsimhan}},
  \bibinfo {author} {\bibfnamefont {B.}~\bibnamefont {Gouveia}}, \bibinfo
  {author} {\bibfnamefont {S.}~\bibnamefont {Kumar}}, \bibinfo {author}
  {\bibfnamefont {E.~S.~G.}\ \bibnamefont {Shaqfeh}}, \ and\ \bibinfo {author}
  {\bibfnamefont {S.~J.}\ \bibnamefont {Muller}},\ }\href
  {http://dx.doi.org/10.1039/C5SM03004H} {\bibfield  {journal} {\bibinfo
  {journal} {Soft Matter}\ }\textbf {\bibinfo {volume} {12}},\ \bibinfo {pages}
  {3787} (\bibinfo {year} {2016})}\BibitemShut {NoStop}%
\bibitem [{\citenamefont {Olla}(2000)}]{olla-pa-2000}%
  \BibitemOpen
  \bibfield  {author} {\bibinfo {author} {\bibfnamefont {P.}~\bibnamefont
  {Olla}},\ }\href {\doibase https://doi.org/10.1016/S0378-4371(99)00563-4}
  {\bibfield  {journal} {\bibinfo  {journal} {Physica A}\ }\textbf {\bibinfo
  {volume} {278}},\ \bibinfo {pages} {87} (\bibinfo {year} {2000})}\BibitemShut
  {NoStop}%
\bibitem [{\citenamefont {Vlahovska}(2016)}]{vlahovska-2016}%
  \BibitemOpen
  \bibfield  {author} {\bibinfo {author} {\bibfnamefont {P.~M.}\ \bibnamefont
  {Vlahovska}},\ }in\ \href {http://dx.doi.org/10.1039/9781782628491-00313}
  {\emph {\bibinfo {booktitle} {Fluid-Structure Interactions in
  Low-Reynolds-Number Flows}}}\ (\bibinfo  {publisher} {The Royal Society of
  Chemistry},\ \bibinfo {year} {2016})\ pp.\ \bibinfo {pages}
  {313--346}\BibitemShut {NoStop}%
\bibitem [{\citenamefont {Lebedev}\ \emph {et~al.}(2007)\citenamefont
  {Lebedev}, \citenamefont {Turitsyn},\ and\ \citenamefont
  {Vergeles}}]{lebedev-prl-2007}%
  \BibitemOpen
  \bibfield  {author} {\bibinfo {author} {\bibfnamefont {V.~V.}\ \bibnamefont
  {Lebedev}}, \bibinfo {author} {\bibfnamefont {K.~S.}\ \bibnamefont
  {Turitsyn}}, \ and\ \bibinfo {author} {\bibfnamefont {S.~S.}\ \bibnamefont
  {Vergeles}},\ }\href {https://dx.doi.org/10.1103/PhysRevLett.99.218101}
  {\bibfield  {journal} {\bibinfo  {journal} {Phys. Rev. Lett.}\ }\textbf
  {\bibinfo {volume} {99}},\ \bibinfo {pages} {218101} (\bibinfo {year}
  {2007})}\BibitemShut {NoStop}%
\bibitem [{\citenamefont {Rustom}\ \emph {et~al.}(2004)\citenamefont {Rustom},
  \citenamefont {Saffrich}, \citenamefont {Markovic}, \citenamefont {Walther},\
  and\ \citenamefont {Gerdes}}]{rustom-science-2004}%
  \BibitemOpen
  \bibfield  {author} {\bibinfo {author} {\bibfnamefont {A.}~\bibnamefont
  {Rustom}}, \bibinfo {author} {\bibfnamefont {R.}~\bibnamefont {Saffrich}},
  \bibinfo {author} {\bibfnamefont {I.}~\bibnamefont {Markovic}}, \bibinfo
  {author} {\bibfnamefont {P.}~\bibnamefont {Walther}}, \ and\ \bibinfo
  {author} {\bibfnamefont {H.-H.}\ \bibnamefont {Gerdes}},\ }\href
  {https://dx.doi.org/10.1126/science.1093133} {\bibfield  {journal} {\bibinfo
  {journal} {Science}\ }\textbf {\bibinfo {volume} {303}},\ \bibinfo {pages}
  {1007} (\bibinfo {year} {2004})}\BibitemShut {NoStop}%
\bibitem [{\citenamefont {Dai}\ and\ \citenamefont
  {Sheetz}(1995{\natexlab{a}})}]{dai-bpj-1995}%
  \BibitemOpen
  \bibfield  {author} {\bibinfo {author} {\bibfnamefont {J.}~\bibnamefont
  {Dai}}\ and\ \bibinfo {author} {\bibfnamefont {M.~P.}\ \bibnamefont
  {Sheetz}},\ }\href {https://doi.org/10.1016/S0006-3495(95)80274-2} {\bibfield
   {journal} {\bibinfo  {journal} {Biophys. J.}\ }\textbf {\bibinfo {volume}
  {68}},\ \bibinfo {pages} {988} (\bibinfo {year}
  {1995}{\natexlab{a}})}\BibitemShut {NoStop}%
\bibitem [{\citenamefont {Cuvelier}\ \emph {et~al.}(2005)\citenamefont
  {Cuvelier}, \citenamefont {Derényi}, \citenamefont {Bassereau},\ and\
  \citenamefont {Nassoy}}]{cuvelier-bpj-2005}%
  \BibitemOpen
  \bibfield  {author} {\bibinfo {author} {\bibfnamefont {D.}~\bibnamefont
  {Cuvelier}}, \bibinfo {author} {\bibfnamefont {I.}~\bibnamefont {Derényi}},
  \bibinfo {author} {\bibfnamefont {P.}~\bibnamefont {Bassereau}}, \ and\
  \bibinfo {author} {\bibfnamefont {P.}~\bibnamefont {Nassoy}},\ }\href
  {https://doi.org/10.1529/biophysj.104.056473} {\bibfield  {journal} {\bibinfo
   {journal} {Biophys. J.}\ }\textbf {\bibinfo {volume} {88}},\ \bibinfo
  {pages} {2714} (\bibinfo {year} {2005})}\BibitemShut {NoStop}%
\bibitem [{\citenamefont {Roux}\ \emph {et~al.}(2002)\citenamefont {Roux},
  \citenamefont {Cappello}, \citenamefont {Cartaud}, \citenamefont {Prost},
  \citenamefont {Goud},\ and\ \citenamefont {Bassereau}}]{roux-pnas-2002}%
  \BibitemOpen
  \bibfield  {author} {\bibinfo {author} {\bibfnamefont {A.}~\bibnamefont
  {Roux}}, \bibinfo {author} {\bibfnamefont {G.}~\bibnamefont {Cappello}},
  \bibinfo {author} {\bibfnamefont {J.}~\bibnamefont {Cartaud}}, \bibinfo
  {author} {\bibfnamefont {J.}~\bibnamefont {Prost}}, \bibinfo {author}
  {\bibfnamefont {B.}~\bibnamefont {Goud}}, \ and\ \bibinfo {author}
  {\bibfnamefont {P.}~\bibnamefont {Bassereau}},\ }\href
  {https://doi.org/10.1073/pnas.082107299} {\bibfield  {journal} {\bibinfo
  {journal} {Proc. Natl. Acad. Sci. U.S.A.}\ }\textbf {\bibinfo {volume}
  {99}},\ \bibinfo {pages} {5394} (\bibinfo {year} {2002})}\BibitemShut
  {NoStop}%
\bibitem [{\citenamefont {Evans}\ and\ \citenamefont
  {Yeung}(1994)}]{evans-yeung-cpl-1994}%
  \BibitemOpen
  \bibfield  {author} {\bibinfo {author} {\bibfnamefont {E.~A.}\ \bibnamefont
  {Evans}}\ and\ \bibinfo {author} {\bibfnamefont {A.}~\bibnamefont {Yeung}},\
  }\href {https://doi.org/10.1016/0009-3084(94)90173-2} {\bibfield  {journal}
  {\bibinfo  {journal} {Chem. Phys. Lipids}\ }\textbf {\bibinfo {volume}
  {73}},\ \bibinfo {pages} {39} (\bibinfo {year} {1994})}\BibitemShut {NoStop}%
\bibitem [{\citenamefont {Dai}\ and\ \citenamefont
  {Sheetz}(1995{\natexlab{b}})}]{dai-cell-1995}%
  \BibitemOpen
  \bibfield  {author} {\bibinfo {author} {\bibfnamefont {J.}~\bibnamefont
  {Dai}}\ and\ \bibinfo {author} {\bibfnamefont {M.~P.}\ \bibnamefont
  {Sheetz}},\ }\href {https://doi.org/10.1016/0092-8674(95)90182-5} {\bibfield
  {journal} {\bibinfo  {journal} {Cell}\ }\textbf {\bibinfo {volume} {83}},\
  \bibinfo {pages} {693} (\bibinfo {year} {1995}{\natexlab{b}})}\BibitemShut
  {NoStop}%
\bibitem [{\citenamefont {Monnier}\ \emph {et~al.}(2010)\citenamefont
  {Monnier}, \citenamefont {Rochal}, \citenamefont {Parmeggiani},\ and\
  \citenamefont {Lorman}}]{monnier-prl-2010}%
  \BibitemOpen
  \bibfield  {author} {\bibinfo {author} {\bibfnamefont {S.}~\bibnamefont
  {Monnier}}, \bibinfo {author} {\bibfnamefont {S.~B.}\ \bibnamefont {Rochal}},
  \bibinfo {author} {\bibfnamefont {A.}~\bibnamefont {Parmeggiani}}, \ and\
  \bibinfo {author} {\bibfnamefont {V.~L.}\ \bibnamefont {Lorman}},\ }\href
  {https://dx.doi.org/10.1103/PhysRevLett.105.028102} {\bibfield  {journal}
  {\bibinfo  {journal} {Phys. Rev. Lett.}\ }\textbf {\bibinfo {volume} {105}},\
  \bibinfo {pages} {028102} (\bibinfo {year} {2010})}\BibitemShut {NoStop}%
\bibitem [{\citenamefont {Rahimi}\ \emph {et~al.}(2013)\citenamefont {Rahimi},
  \citenamefont {DeSimone},\ and\ \citenamefont
  {Arroyo}}]{rahimi-soft-matter-2013}%
  \BibitemOpen
  \bibfield  {author} {\bibinfo {author} {\bibfnamefont {M.}~\bibnamefont
  {Rahimi}}, \bibinfo {author} {\bibfnamefont {A.}~\bibnamefont {DeSimone}}, \
  and\ \bibinfo {author} {\bibfnamefont {M.}~\bibnamefont {Arroyo}},\ }\href
  {https://dx.doi.org/10.1039/C3SM51748A} {\bibfield  {journal} {\bibinfo
  {journal} {Soft Matter}\ }\textbf {\bibinfo {volume} {9}},\ \bibinfo {pages}
  {11033} (\bibinfo {year} {2013})}\BibitemShut {NoStop}%
\bibitem [{\citenamefont {Sahu}\ \emph {et~al.}(2020)\citenamefont {Sahu},
  \citenamefont {Omar}, \citenamefont {Sauer},\ and\ \citenamefont
  {Mandadapu}}]{sahu-mandadapu-ale-i}%
  \BibitemOpen
  \bibfield  {author} {\bibinfo {author} {\bibfnamefont {A.}~\bibnamefont
  {Sahu}}, \bibinfo {author} {\bibfnamefont {Y.~A.~D.}\ \bibnamefont {Omar}},
  \bibinfo {author} {\bibfnamefont {R.~A.}\ \bibnamefont {Sauer}}, \ and\
  \bibinfo {author} {\bibfnamefont {K.~K.}\ \bibnamefont {Mandadapu}},\ }\href
  {\doibase 10.1016/j.jcp.2020.109253} {\bibfield  {journal} {\bibinfo
  {journal} {J. Comput. Phys.}\ }\textbf {\bibinfo {volume} {407}},\ \bibinfo
  {pages} {109253} (\bibinfo {year} {2020})},\ \Eprint
  {http://arxiv.org/abs/1812.05086} {arXiv:1812.05086} \BibitemShut {NoStop}%
\bibitem [{\citenamefont {Bar-Ziv}\ \emph {et~al.}(1999)\citenamefont
  {Bar-Ziv}, \citenamefont {Tlusty}, \citenamefont {Moses}, \citenamefont
  {Safran},\ and\ \citenamefont {Bershadsky}}]{bar-ziv-pnas-1999}%
  \BibitemOpen
  \bibfield  {author} {\bibinfo {author} {\bibfnamefont {R.}~\bibnamefont
  {Bar-Ziv}}, \bibinfo {author} {\bibfnamefont {T.}~\bibnamefont {Tlusty}},
  \bibinfo {author} {\bibfnamefont {E.}~\bibnamefont {Moses}}, \bibinfo
  {author} {\bibfnamefont {S.~A.}\ \bibnamefont {Safran}}, \ and\ \bibinfo
  {author} {\bibfnamefont {A.}~\bibnamefont {Bershadsky}},\ }\href
  {htts://dx.doi.org/10.1073/pnas.96.18.10140} {\bibfield  {journal} {\bibinfo
  {journal} {Proc. Natl. Acad. Sci. U.S.A.}\ }\textbf {\bibinfo {volume}
  {96}},\ \bibinfo {pages} {10140} (\bibinfo {year} {1999})}\BibitemShut
  {NoStop}%
\bibitem [{\citenamefont {Upadhyaya}\ and\ \citenamefont
  {Sheetz}(2004)}]{upadhyaya-bpj-2004}%
  \BibitemOpen
  \bibfield  {author} {\bibinfo {author} {\bibfnamefont {A.}~\bibnamefont
  {Upadhyaya}}\ and\ \bibinfo {author} {\bibfnamefont {M.~P.}\ \bibnamefont
  {Sheetz}},\ }\href {https://doi.org/10.1016/S0006-3495(04)74343-X} {\bibfield
   {journal} {\bibinfo  {journal} {Biophys. J.}\ }\textbf {\bibinfo {volume}
  {86}},\ \bibinfo {pages} {2923} (\bibinfo {year} {2004})}\BibitemShut
  {NoStop}%
\bibitem [{\citenamefont {Kaether}\ \emph {et~al.}(2000)\citenamefont
  {Kaether}, \citenamefont {Skehel},\ and\ \citenamefont
  {Dotti}}]{kaether-mbc-2000}%
  \BibitemOpen
  \bibfield  {author} {\bibinfo {author} {\bibfnamefont {C.}~\bibnamefont
  {Kaether}}, \bibinfo {author} {\bibfnamefont {P.}~\bibnamefont {Skehel}}, \
  and\ \bibinfo {author} {\bibfnamefont {C.~G.}\ \bibnamefont {Dotti}},\ }\href
  {https://doi.org/10.1091/mbc.11.4.1213} {\bibfield  {journal} {\bibinfo
  {journal} {Mol. Biol. Cell}\ }\textbf {\bibinfo {volume} {11}},\ \bibinfo
  {pages} {1213} (\bibinfo {year} {2000})}\BibitemShut {NoStop}%
\bibitem [{\citenamefont {Leduc}\ \emph {et~al.}(2004)\citenamefont {Leduc},
  \citenamefont {Camp{\`a}s}, \citenamefont {Zeldovich}, \citenamefont {Roux},
  \citenamefont {Jolimaitre}, \citenamefont {Bourel-Bonnet}, \citenamefont
  {Goud}, \citenamefont {Joanny}, \citenamefont {Bassereau},\ and\
  \citenamefont {Prost}}]{leduc-pnas-2004}%
  \BibitemOpen
  \bibfield  {author} {\bibinfo {author} {\bibfnamefont {C.}~\bibnamefont
  {Leduc}}, \bibinfo {author} {\bibfnamefont {O.}~\bibnamefont {Camp{\`a}s}},
  \bibinfo {author} {\bibfnamefont {K.~B.}\ \bibnamefont {Zeldovich}}, \bibinfo
  {author} {\bibfnamefont {A.}~\bibnamefont {Roux}}, \bibinfo {author}
  {\bibfnamefont {P.}~\bibnamefont {Jolimaitre}}, \bibinfo {author}
  {\bibfnamefont {L.}~\bibnamefont {Bourel-Bonnet}}, \bibinfo {author}
  {\bibfnamefont {B.}~\bibnamefont {Goud}}, \bibinfo {author} {\bibfnamefont
  {J.-F.}\ \bibnamefont {Joanny}}, \bibinfo {author} {\bibfnamefont
  {P.}~\bibnamefont {Bassereau}}, \ and\ \bibinfo {author} {\bibfnamefont
  {J.}~\bibnamefont {Prost}},\ }\href
  {https://dx.doi.org/10.1073/pnas.0406598101} {\bibfield  {journal} {\bibinfo
  {journal} {Proc. Natl. Acad. Sci. U.S.A.}\ }\textbf {\bibinfo {volume}
  {101}},\ \bibinfo {pages} {17096} (\bibinfo {year} {2004})}\BibitemShut
  {NoStop}%
\bibitem [{\citenamefont {Tchoufag}\ \emph {et~al.}(tion)\citenamefont
  {Tchoufag}, \citenamefont {Sahu},\ and\ \citenamefont
  {Mandadapu}}]{tchoufag-convective}%
  \BibitemOpen
  \bibfield  {author} {\bibinfo {author} {\bibfnamefont {J.}~\bibnamefont
  {Tchoufag}}, \bibinfo {author} {\bibfnamefont {A.}~\bibnamefont {Sahu}}, \
  and\ \bibinfo {author} {\bibfnamefont {K.~K.}\ \bibnamefont {Mandadapu}},\
  }\href@noop {} {\  (\bibinfo {year} {In preparation})}\BibitemShut {NoStop}%
\bibitem [{\citenamefont {Kreyszig}(1968)}]{kreyszig}%
  \BibitemOpen
  \bibfield  {author} {\bibinfo {author} {\bibfnamefont {E.}~\bibnamefont
  {Kreyszig}},\ }\href@noop {} {\emph {\bibinfo {title} {Introduction to
  Differential Geometry and Riemannian Geometry}}}\ (\bibinfo  {publisher}
  {University of Toronto Press},\ \bibinfo {address} {Toronto},\ \bibinfo
  {year} {1968})\BibitemShut {NoStop}%
\bibitem [{\citenamefont {Willmore}(1996)}]{willmore}%
  \BibitemOpen
  \bibfield  {author} {\bibinfo {author} {\bibfnamefont {T.~J.}\ \bibnamefont
  {Willmore}},\ }\href@noop {} {\emph {\bibinfo {title} {Riemannian
  Geometry}}}\ (\bibinfo  {publisher} {Oxford University Press},\ \bibinfo
  {address} {Oxford},\ \bibinfo {year} {1996})\BibitemShut {NoStop}%
\bibitem [{\citenamefont {Naghdi}(1973)}]{naghdi-1973-theory}%
  \BibitemOpen
  \bibfield  {author} {\bibinfo {author} {\bibfnamefont {P.~M.}\ \bibnamefont
  {Naghdi}},\ }\href@noop {} {\emph {\bibinfo {title} {The Theory of Shells and
  Plates}}}\ (\bibinfo  {publisher} {Springer Berlin Heidelberg},\ \bibinfo
  {year} {1973})\ pp.\ \bibinfo {pages} {425--640}\BibitemShut {NoStop}%
\bibitem [{\citenamefont {Parkkila}\ \emph {et~al.}(2018)\citenamefont
  {Parkkila}, \citenamefont {Elderdfi}, \citenamefont {Bunker},\ and\
  \citenamefont {Viitala}}]{parkkila-langmuir-2018}%
  \BibitemOpen
  \bibfield  {author} {\bibinfo {author} {\bibfnamefont {P.}~\bibnamefont
  {Parkkila}}, \bibinfo {author} {\bibfnamefont {M.}~\bibnamefont {Elderdfi}},
  \bibinfo {author} {\bibfnamefont {A.}~\bibnamefont {Bunker}}, \ and\ \bibinfo
  {author} {\bibfnamefont {T.}~\bibnamefont {Viitala}},\ }\href
  {https://dx.doi.org/10.1021/acs.langmuir.8b01259} {\bibfield  {journal}
  {\bibinfo  {journal} {Langmuir}\ }\textbf {\bibinfo {volume} {34}},\ \bibinfo
  {pages} {8081} (\bibinfo {year} {2018})}\BibitemShut {NoStop}%
\end{thebibliography}%

\end{document}